%% file: mainams.tex
\documentclass[10pt,a4papper]{amsart}
\usepackage[english]{babel}
\usepackage{graphicx,graphics}
\usepackage[section]{placeins}
\usepackage{hhline,xspace,exscale}
\usepackage{amsmath,amsfonts,amssymb}
\usepackage{latexsym}
\usepackage{eufrak}
\include{lucmath}% definitions of styles and commands
\begin{document}
%----------------------------------------------------------------
\title[Structure of solutions of the Skyrme model on
three-sphere]{Structure of solutions of the Skyrme model on
three-sphere\\ Numerical results}
\author{{\L}ukasz Bratek}

\subjclass{}
\date{\today}
%--------------------------- ABSTRACT ----------------------------
\begin{abstract}
The hedgehog Skyrme model on three-sphere admits very rich
spectrum of solitonic solutions which can be encompassed by a
strikingly simple scheme. The main result of this paper is the
statement of the tripartite structure of solutions of the model
and the discovery in what configurations these solutions appear.
The model contains features of more complicated models in General
Relativity and as such can give insight into them.
\end{abstract}
%---------------------------- MAIN -------------------------------
\maketitle {\small \noindent \textsf{Department of Physics,
Mathematics \& Computer Science \\ \noindent Jagellonian
University, Reymonta 4, Cracow, Poland.\\ \noindent
email:\textmd{\
bratek@th.if.uj.edu.pl}\\
\noindent Submitted to Nonlinearity}}
\medskip
\input{introduction}
\input{hedgehogmodel}
\input{triplestructure}

\input{wholestructure}
\input{acknwld01}
\appendix \label{apdx}
\input{appendix}
\medskip\medskip\medskip\medskip
%GATHER{Xbib.bib}
\bibliographystyle{amsplain}
\bibliography{xbib}
% ----------------------------------------------------------------
\end{document}

%% file: lucmath.tex
\newtheorem{thm}{Theorem}[section]
\newtheorem{hpt}{Hypothesis}[section]

\newtheorem{lem}[thm]{Lemma}

\theoremstyle{definition}
\newtheorem{dfn}[thm]{Definition}
\theoremstyle{remark}

\numberwithin{equation}{section}
\newcommand{\s}{\mathcal{S}^3}
\newcommand{\sk}[1]{\textsf {\emph{S}}_{\scriptscriptstyle{{#1}}}}
\newcommand{\as}[1]{\overline{\textsf{\emph{S}}}_{\scriptscriptstyle{{#1}}}}
\newcommand{\ha}[1]{\textsf{\emph{H}}_{\scriptscriptstyle{{#1}}}}
\newcommand{\ah}[1]{\overline{\textsf{\emph{H}}}_{\scriptscriptstyle{{#1}}}}
\newcommand{\sks}[1]{\textsf {\emph{s}}_{\scriptscriptstyle{{#1}}}}

\newcommand{\has}[1]{\textsf{\emph{h}}_{\scriptscriptstyle{{#1}}}}
\newcommand{\ahs}[1]{\overline{\textsf{\emph{h}}}_{\scriptscriptstyle{{#1}}}}

\renewcommand{\c}[2]{\mathfrak{F}_{#1}^{#2}}

\newcommand{\q}{\textsf{\emph{Q}}}

\newcommand{\half}{\textstyle \frac{1}{2}\displaystyle }
\newcommand{\quarter}{\textstyle \frac{1}{4}\displaystyle }
\newcommand{\oct}{\textstyle \frac{1}{8}\displaystyle }

\newcommand{\olarge}{\mathcal{O}}
\newcommand{\osmall}{\scriptscriptstyle{\mathcal{O}}\displaystyle}

\newcommand{\tr}{\textsf{Tr\ }}
\newcommand{\f}{\textsf{\emph{f}}_\pi}
\newcommand{\e}{\textsf{\emph{e}}}
\newcommand{\ud }{\mathrm{d}}

\newcommand{\bizon}{Bizo\'{n}\xspace }

\newcommand{\dv}[2]{\begin{tabular}[h]{l}\!\!\!$#1$\!\!\! \\ \!\!\!$#2$\!\!\! \end{tabular}}

%% file: introduction.tex
\section{
Introduction to the Skyrme model on $\s$ and motivation}

In this paper I discuss static solutions of the Skyrme model on
the three-sphere of radius $R$ as a physical space. In a sense, to
be explained later this model includes the Skyrme model on flat
space $\mathbb{R}^3$ in the limit $R\rightarrow \infty $. The
model contains two scales of length which constitute a free
dimensionless parameter whose value is crucial to the existence of
solutions of a given type. There are known several models in which
similar behaviour is observed (\emph{e.g.} \cite{bizchmaj}) and
because of this it deserves investigation. The Skyrme model on
$\s$ is the simplest nonlinear realization of such models which
admits very rich structure of solitonic solutions.

The Skyrme model on flat space was originally introduced as
minimal nonlinear realization of an effective meson field theory
where low energy baryons emerged as solitonic fields. Then
nucleons were introduced as topological excitations of these
fields. The basic chiral field in that model is $SU(2)$ valued
scalar field $U(x)$ which is related to the singlet meson field
$\sigma $ and the triplet pion field $\vec{\pi }$ by
$$U(\vec{x})=
\frac{1}{\f }(\sigma(\vec{x}) +i\vec{\sigma}\circ \vec{\pi
}(\vec{x})),\qquad (\sigma)^2 +(\vec{\pi })^2=\f ^2.$$ The model
is defined by the Lagrange density
\begin{equation}
\mathcal{L}=\frac{1}{4}\f^2\tr(L _\mu L^\mu)
+\frac{1}{32\e^2}\tr([L_\mu ,L_\nu][L^\mu ,L^\nu])
\label{eq:lagdens},\end{equation} where $$L_\mu=U^+\partial _\mu
U$$ is $su(2)$ valued topological current. The model is
characterized by two constants $\f$ - pion decay constant and $\e$
determining the strength of the fourth order term which was
introduced by Skyrme in \cite{skyrme3} in order to ensure
existence of solitonic solutions, but it can be introduced in
quite different way - on geometrical level which I shall sketch
now.

The Lagrange density (\ref{eq:lagdens}) has  a nice geometrical
origin. An  $m\textrm{-parameter}$ Lie group $\mathcal{G}$ valued
scalar field $U$ is a map $\mathcal{M}\ni p\rightarrow U(p) \in
\mathcal{G} $. If $x=x(p)$ is a local coordinate system on
$\mathcal{M}$ and $\xi$ are co-ordinates of the Lie group about
the identity of the group, then the field induces a map
$\mathbb{R}^n\ni x\rightarrow \xi(x) \in \mathbb{R}^m$ and
respective mapping of the tangent bundle $T\mathcal{M}$ to the Lie
algebra $T_{\mathbf{e}}\mathcal{G}$ - the tangent space to the Lie
group $\mathcal{G}$ at the identity. Each one-parameter subgroup
$U(t)$ of $\mathcal{G}$ such that $U(t_1+t_2)=U(t_1)U(t_2)$ and
$U(0)=\mathbf{e}$ defines a smooth curve $U(t)$ with tangent
vector $\dot{U}(t)=U(t)u \in T_{U(t)}\mathcal{G}$ where $u \in
T_\mathbf{e}\mathcal{G}$ is tangent to $\mathbf{e}$, so that
$U^{-1}\dot{U}(t)$ is an element of $T_{\mathbf{e}}\mathcal{G}$.
This fact may  be used to define the \emph{repere mobile} of the
Lie algebra at the identity by
$$\mathbf{a}_i(\xi):=U^{-1}\partial_iU(\xi).$$ By that, with each
$v\in T_p\mathcal{M}$ we can associate an element of the Lie
algebra  $$T_p\mathcal{M}\ni v\rightarrow \mathbf{a}_i(\xi
(x))\frac{\partial \xi ^i }{\partial x^\mu}v^\mu(x)\in
T_{\mathbf{e}}\mathcal{G},$$ which defines a Lie algebra valued
one-form $\Omega=-U^{-1}\partial_\mu U\ud x^\mu$ on $\mathcal{M}$
which simultaneously is a matrix of connection one-forms whose
curvature identically vanishes
$d\Omega-\Omega\wedge\Omega\equiv0$. From the connection we
construct secondary quantities $\Theta:=\ud \Omega$ and, since
$\mathcal{M}$ is a Riemannian manifold, $\ast \Omega$ and $\ast
\Theta$ (where $\ast$ is the Hodge operator). Now we can build
invariants being full scalar products $(\ast \Omega,\Omega)$ and
$(\ast \Theta,\Theta)$, taking the first as the sigma term and the
second as the self-interaction term for our field such that,
imposing on $\mu$, $\lambda$, and $U$ physical dimensions, we
arrive at (\ref{eq:lagdens}) starting from
$$\int \limits_{\mathcal{M}}\mathcal{L}(U,\partial{U})\sqrt{-g}\ud^4x
:=\int \limits_{\mathcal{M}}\mu\tr(\ast\Omega
\wedge\Omega)+\lambda\tr(\ast \Theta\wedge\Theta).$$ This equation
defines, in generally covariant manner, the lagrange density
$\mathcal{L}$ and generalizes the flat Skyrme model to an
arbitrary curved space-time. Another point of view, though only
for static field configurations, was presented by Krush in
\cite{krush} on the level of density of energy, and can be recast
into the following scheme. Taking, as a physical quantity, the
strain tensor $D_{ik}=-\half \tr
(U^{-1}\partial_iUU^{-1}\partial_kU)$ being the Killing form for
the Lie algebra of the compact matrix group $SU(2)$, we use
invariants originating from the secular equation
$\det(1+tD)=c_kt^k$, which serves as generating polynomial in $t$.
The invariants are respectively $c_1=\tr D$, $c_2=\half (\tr
D)^2-\half \tr D^2 $, which in particular reproduces the sigma and
self-interaction terms in ($\ref{eq:endens}$), and additionally
$c_3=\det(D)$ which is proportional to the density of topological
charge. This approach generalizes the energy density for static
solutions to an arbitrary curved three-geometry. (Aside remark:
such invariants can be easily derived if one notices that
$\det(A)=\exp(\tr\log(A))$ what is obvious for diagonalizable
matrices and, by virtue of continuity, must hold in general.
Putting $A=E+tD$, expanding in the variable $t$, and remembering
that for matrices $D$ of rank $n$ the Taylor series expansion in
$t$ must terminate at $t^n$, we get all invariants for $D$).

The lagrange density (\ref{eq:lagdens}) leads to the action for a
field $U$:
$$\mathcal{S}[U]=\int_{\mathcal{M}}\mathcal{L}(U,\partial{U})\sqrt{-g}\ud^4x.$$

Here we immerse the Skyrme model within the simplest possible
curved  space-time which, from now on, we assume to have the
structure of the Cartesian product
$\mathcal{M}=\mathbb{R}\times\s_R$ ($\s_R$ is the three-sphere of
radius $R$) which is mapped into a subset of $\mathbb{R}^4$ by the
map $\mathbb{R}\times\s_R\ni p\rightarrow
\Upsilon(p)=(t,\psi,\vartheta,\varphi)$. In this map the line
element takes the form
$$\ud s^2=\ud t^2-R^2\ud \psi^2- R^2\sin^2{\psi}\big(\ud
\vartheta^2+\sin^2{\vartheta}\ud \varphi^2\big)$$ and,
disregarding singularities of the map, we assume that
$t\in(-\infty, +\infty)$, $\psi \in [0,\pi]$, $\vartheta \in
[0,\pi]$ and $\varphi \in [0,2\pi)$. For that space-time is
invariant with respect to the group of translations in time,  the
first theorem of Noether assures for solutions the existence of a
constant of motion - the energy integral. Moreover solutions which
are static, and we limit ourself in this paper considering only
such one, are critical points of the integral
$(g^{ik}_{\s}:=-\frac{1}{R^2}g^{ik})$
\begin{equation}E=R^3\int_{\s}T_{00}\sqrt{g_{\s}}\ud^3{\xi},
\label{eq:energint} \end{equation} where $T_{00}$ is $00$
component of the metrical energy-momentum tensor, which for static
solutions takes the form (then $T_{00}=-\mathcal{L}$)
\begin{equation}T_{00}=\frac{1}{4}\frac{\f^2}{R^2}g_{\s}^{ik}\tr(L _i L_k)+
\frac{1}{16\e^2R^4}g_{\s}^{jk}g_{\s}^{il}\tr(L_iL_j[L_k,L_l]).\label{eq:endens}
\end{equation}
The integral \ref{eq:energint} measures  the energy of a map from
the configuration space $\s$ to the $SU(2)$ group which, as a
Riemannian manifold, is equivalent to $\s$. So that we examine
energies for mappings from $\s$ into $\s$.

The Euler-Lagrange equations for the Skyrme model on
 $\s$, defined above by the energy integral
(\ref{eq:energint}), lead to a system of three nonlinear partial
differential equations of second order for three unknown
functions. To bypass this outstanding example of human endeavour,
exploiting the invariance of (\ref{eq:energint}) under the action
of $SO(3)$ group of rotations, we restrict ourselves by
considering only those fields which are $SO(3)$ invariant.
 According to the theorem given by Faddeev in \cite{fadeev} an
extremum of a functional in the domain of invariant fields under
the group of invariance of the functional is the extremum of the
functional in the domain of all fields. The hedgehog Ansatz
provides us with a subgroup of fields which are $SO(3)$ invariant.
The hedgehog field, which in original Skyrme model describes
baryons, is defined by
\begin{equation}
U(\psi,\varphi,\vartheta)=e^{i\vec{\sigma }\circ
\vec{n}(\varphi,\vartheta)F(\psi)}, \label{eq:ansatze}
\end{equation} and in
language of Skyrme the pion field $\vec{\pi}$ is identified with
the direction $\vec{n}(\varphi,\vartheta)$ in isospin space  and
$F(\psi )$ measures its amplitude. The form of this Ansatz, of
course, comes from the canonical parameterization of $SU(2)$ group
with addition of implicitly  assumed equivariance of the map
$\s\rightarrow\s$ \emph{i.e.} $(\psi,\vartheta,\varphi)\rightarrow
(\Psi,\Theta,\Phi)$ such that
$\Psi=F(\psi),\Theta\equiv\vartheta,\Phi\equiv\phi$. This is in
accordance with assumptions made in \cite{biz} and by that we get
a natural field-theoretical generalization of the model of
harmonic maps between three-spheres of \bizon . Our model contains
a free parameter $R$ which determines the curvature of physical
space and, as we shall see, has crucial effect on the number of
solutions.
% which cannot exist if the radius is too small.
This is a toy model which mimics critical phenomena in formation
of solutions in the Einstein-Skyrme model \cite{bizchmaj} (but not
only) and in principle it was the main motivation of the present
work.

The hedgehog Ansatz leads to solutions which in the original
Skyrme model were called skyrmions. In general we call skyrmions
all classical static solutions of the model which minimize energy
within a given sector of the baryon number $B$. Equations of
motion for the Skyrme fields are the Euler-Lagrange equations for
the energy functional and, as such, the solutions may be not
minima but saddle points.

In the case of the Skyrme model on flat space, if $U$ tends to a
certain limiting matrix at spatial infinity independent of any
direction, we can define $U(\infty )=\lim_{|\vec{x}|\rightarrow
\infty}U(\vec{x}) $. If so, we can attach to $\mathbb{R}^3$ the
point at infinity and by that complete the space to a compact one.
Next we can identify points from $\mathbb{R}^3\cup \{\infty \}$
with points from $\s $ by a suitable mapping - for instance the
stereographic projection. A mapping from $\mathbb{R}^3\cup
\{\infty \}$ to $\s $ and further from
$\mathbb{T}\times(\mathbb{R}^3\cup \{\infty \})$ to
$\mathbb{T}\times \s$ is not an isometry and produces space-time
with metric tensor whose curvature is not vanishing. It means that
the Skyrme model on flat space is not metrically equivalent to the
transformed one but topologically is. By means of that topological
equivalence we can examine certain topological aspects of the flat
Skyrme model in compact space such as $\s$ what partly explains my
work on the Skyrme model on $\s $. But we can look at this from
quite another point of view. Physically a space-time endowed with
non flat metrical tensor stands for gravity but the space-time
from this point of view is then only a scene for skyrmions;
space-time influences their motion but they do not influence  its
geometry. It may of course be so only under certain conditions
\emph{e.g.} if the presence of skyrmions only slightly disturbs
metric tensor (like gravitational waves far away from their
source). Then equations of motion for skyrmions decouple from
Einstein's equations and may be solved separately.

My aim here is not investigation of the full set of equations of
the Einstein-Skyrme model (see \cite{bizchmaj}). I am looking for
some aspects of nonlinear field theory coupled to gravity
independent on the specific items of the space-time. The Skyrme
model on $\s$ is a toy model and as such aspires only to give
insight into these unexplored areas of nonlinearity under
gravitation. Here we treat the model only as an example of
nonlinear field theory.

%% file: hedgehogmodel.tex
\section{The hedgehog Skyrme model on $\s$}

In this paper I consider only static solutions of the Skyrme model
on three-sphere which are critical points of the energy functional
(\ref{eq:energint}). Using the hedgehog Ansatz (\ref{eq:ansatze})
we get
$$E[F]=\int \limits_{0}^\pi 4\pi \sin^2{\psi}\ud\psi\bigg\{ \frac{R\f^2}{2}\bigg[F'^2+
2\frac{\sin^2{F}}{\sin^2{\psi}}
\bigg]+\frac{1}{R\e^2}\frac{\sin^2{F}}{\sin^2{\psi}}
\bigg[F'^2+\frac{1}{2}\frac{\sin^2{F}}{\sin^2{\psi}}\bigg]\bigg\}.$$
In the limit $\e \rightarrow \infty$ we regain the energy
functional (with the unit of energy $R\f^2/2$) for harmonic maps
between three-spheres which are equivariant with respect to the
action of $SO(3)$ group and which were examined by \bizon  in
\cite{biz}. Thus the Skyrme term may be considered as a
perturbation which  deforms harmonic maps of \bizon.

Two parameters $\e$ and $\f$ (in units in which $c=1$) provide us
with the natural unit of length $L_o:=(\e\f)^{-1}$, determining
characteristic size of a soliton, and with the unit of energy
$E_o=\f \e^{-1}/2$.
The second scale of length $L_g$ is provided by the radius $R$ of the %%@
physical space - the three sphere. Roughly speaking, $R$ artificially %%@
introduces gravity. (In the Einstein-Skyrme model, the second scale is introduced by %%@
the combination $\sqrt{G}/\e$ originating from the full %%@
Einstein-Skyrme action \cite{bizchmaj}). These two scales of
length constitute a dimensionless
parameter $L:=L_g/L_o=\e \f R$ which is a free parameter of the Skyrme %%@
model on three-sphere.  When $\e$ and  $\f $ are nonzero and
fixed, then $L$ has the interpretation of a dimensionless radius
of the three-sphere. In these units the energy %%@
functional takes the form
\begin{equation}E[F]=4\pi L\int \limits_{0}^\pi %%@
\sin^2{\psi}\ud\psi\bigg\{ \bigg[F'^2+2\frac{\sin^2{F}}{\sin^2{\psi}} %%@
\bigg]
+\frac{2}{L^2}\bigg[F'^2+\frac{1}{2}\frac{\sin^2{F}}{\sin^2{\psi}}\bigg]
\frac{\sin^2{F}}{\sin^2{\psi}}\bigg\}. \label{eq:energy}
\end{equation}
Critical points of the functional (\ref{eq:energy}) are solutions
of   the Euler-Lagrange equation
\begin{equation}
(\sin ^2{\psi }+\kappa ^2\sin ^2{F})F'' +\sin{2\psi
}F'+\frac{1}{2}\kappa ^2\sin{2F}(F')^2-\sin{2F}-\frac{1}{2} \kappa
^2 \frac{\sin ^2{F}}{\sin ^2{\psi }}\sin{2F}=0,\quad
\kappa^2:=\frac{2}{L^2}. \label{eq:main}
\end{equation}
This is the quasilinear second order equation for the shape
function $F(\psi )$. In what follows we shall refer to the
parameter $\kappa^2=2/L^2$ as the \textsl{coupling constant}. %%@
The only nontrivial  solution of this equation, which is known
analytically, is the identity map $F(\psi)=\psi$ which has energy
\begin{equation}E=6\pi ^2(L+L^{-1})\geqslant  12\pi^2;\label{eq:energid}
\end{equation} where, depending on whether $L>1$ or $L<1$,
the harmonic or the Skyrme term prevails.

We are interested only in solutions for which both $F(0)$ and
$F(\pi)$ are integer multiples of $\pi$. Another possibility are
the solutions for which $F(0)$ or $F(\pi)$ (or both) are some odd
multiples of $\pi/2$ (see the appendix for appropriate proofs).
These solutions behave like the function $\sqrt{\lambda}\sin
{\ln(\lambda)}$ about $\lambda=0$, oscillating infinitely many
times in the vicinity of poles. From the physical reasons we give
up considering them, since otherwise the energy integral
($\ref{eq:energint}$) would be divergent.

For solutions, for which both $F(0)$ and $F(\pi)$ are integer
multiples of $\pi$, the Bogomolnyi bound holds, and we get it from
(\ref{eq:energy}) by completion into the squares
\begin{equation}
E[F]=\small{4\pi L\int \limits_{0}^\pi \sin^2{\psi}\ud\psi\bigg\{ %%@
\bigg[F'-\frac{1}{L} \frac{\sin^2{F}}{\sin^2{\psi}} \bigg]^2
+\frac{2}{L^2}\frac{\sin^2{F}}{\sin^2{\psi}}
\bigg[\big(F'-L\big)^2+3LF'\bigg]\bigg\}\geqslant 24\pi \int
\limits_{F(0)}^{F(\pi )} \sin^2{F}\ud F}=12\pi ^2\q.
\label{eq:bog}\end{equation} This bound is quite general and holds
for all $SU(2)$ fields of the model (\ref{eq:energint}). It is
clear that the Bogomolnyi bound is saturated  by the identity map
at $L=1$. By the way the solution which is the counterpart of the
flat one-skyrmion, converges uniformly to the identity when
$L\searrow\sqrt{2}$. The bound $12\pi^2$ is not attained in the
case of flat skyrmions.

By equatorial reflection symmetry of the energy functional
(\ref{eq:energy}), which does not change energy, we can assume,
without loss of generality, that for solutions
$F(\pi)-F(0)\geqslant 0$. For solutions the integer
$$\q=\frac{1}{\pi}\big(F(\pi)-F(0)\big)$$ is
the topological charge which is topological constant of motion and
cannot be changed by continuous deformation of a solution.

Because of symmetries of the functional (\ref{eq:energy}) we can
narrow down the group of solutions. It suffices to take into
account only solutions for which $F(0)=0$ and $F'(0)\geqslant0$,
for if $F(\psi)$ with $F(0)= k\pi$, $k\in \mathbb{Z}$ is a
solution then $F(\psi)-k\pi$ and $-F(\psi)$ are solutions too.
Moreover if $F(\psi)$ is a solution with topological charge $\q$
and with $F(0)=0$, then $G(\psi):=\q\pi -F(\pi -\psi )$ is a
solution with $G(0)=0$ and with the same topological charge $\q$
but with $G'(\psi)=F'(\pi-\psi)$. Hence it suffices to consider
only solutions for which $F'(0)\geqslant F'(\pi)$.

So, to sum up, we can limit ourself, not losing generality, taking
into account only solutions for which $F(0)=0$, $F'(0)\geqslant 0$
and $F'(0)\geqslant F'(\pi)$; of course in special cases $F'(\pi)$
can be negative. Numerics shows that for such solutions
$\q\geqslant 0$.

%% file: triplestructure.tex
\section{Tripartite structure of solutions}\label{chp:howfind}
\input{howfind}
\subsection{Terminology and notation}

Numerics shows that all solutions of the Skyrme model on $\s$ in
general consist of three clearly distinguishable parts: a harmonic
map in the middle and two skyrmions, each with certain topological
charge, attached at the ends of the map. In other words in the
vicinity of poles of the base three-sphere the solutions manifest
their skyrmionic nature and in between they contain a harmonic map
known from the paper of \bizon \cite{biz}. This should be
understood in the following way.

Firstly we consider skyrmionic 'arms' that is the integral curves
of equation (\ref{eq:main}) in the vicinity of poles. We focus our
attention on the north pole. (In the case of the south pole one
proceeds in the analogous way). In order to pass from the Skyrme
model on $\s$ to the ordinary Skyrme model on flat space, we
extract the skyrmionic part from the equation of motion
(\ref{eq:main}) introducing new dimensionless variable $r:=2L\psi$
and defining the function $S(r):=F(\psi(r))$. In a small
neighborhood of the north pole $r \ll 2L$ and consequently
$\sin{\psi}\approx r/2L$. For all finite $r$ we formally get, in
the limit of infinite radius $L$ (the unit of length is fixed),
the equation for the limiting function $S(r)$
\begin{equation}(\quarter r^2+2\sin^2{S})S''+\half
rS'+S'^2\sin{2S} -\quarter \sin{2S}
-\frac{\sin^2{S}}{r^2}\sin{2S}=0\label{eq:skyrmodel}\end{equation}
which is the equation for shape function of the hedgehog skyrmions
on flat three-space \cite{witten}. Here $r$ is the radial
coordinate and the unit of length is
$\e^{-1}\textsf{\emph{F}}^{-1}_\pi$ with
$\textsf{\emph{F}}_\pi=2\f$. The difference in units comes from
the conventions assumed by different authors. In the model,
described by equation (\ref{eq:skyrmodel}), for the reasons
similar to those above, one looks for solutions which interpolate
between $0$ at $r=0$ and $n\pi$ at $r=\infty$. Such solutions have
finite energy and asymptotically behave as $S_n(r)=\alpha_n
r+\olarge(r^3)$ in the vicinity  of $r=0$ and as
$S_n(r)=n\pi+\beta_n/r^2+\osmall(r^{-2})$ at infinity \cite{mat}.
Comparing the first series expansion with (\ref{eq:expansion}) it
is clear that $F'(0)$ for skyrmionic arms must behave like $1/L$
in the limit $L\rightarrow \infty$.  We will denote by $\sks{n}$
such limiting arm, shrunk into the north pole, where $n$ is a
topological charge which simultaneously is the topological charge
we ascribe to the corresponding skyrmionic arm of a solution of
equation (\ref{eq:main}) when $L$ is finite. The skyrmionic arm of
the last solution we will denote by $\sk{n}$ to indicate that it
evolves into $\sks{n}$ as $L\rightarrow \infty$. For the south
pole we proceed in the similar way putting $r=2L(\pi -\psi)$. In
both cases (the north and the south pole) we have to do with
skyrmions if $F'(0)$ or $F'(\pi)$ diverge to $+\infty$ or with
anti-skyrmions if the respective derivatives diverge to $-\infty$
as $L\rightarrow \infty$.

In the second step we consider another case in which for a
solution of equation (\ref{eq:main}) in the limit $L\rightarrow
\infty$ $F'(0)$, $F'(\pi)$ or both are finite. If in this limit
$F'(0)$ (or $F'(\pi)$) is finite then the respective skyrmionic
arm is absent and we have to do with a harmonic map directly
attached to the north pole (the south pole) or with pure harmonic
map if both derivatives are finite.  Harmonic maps appear in our
model since if one drops the Skyrme term, what formally can be
done by taking the limit $L\rightarrow \infty$ in equation
(\ref{eq:main}), we regain the equation for equivariant harmonic
maps between three-spheres. These maps were first found and their
existence proved by \bizon in \cite{biz}. They are solutions of
the equation
\begin{equation}F''(\psi)\sin^2{\psi}+F'(\psi)\sin{2\psi}
-\sin{2F(\psi)}=0.\label{eq:harmap}
\end{equation}
It was proved in \cite{biz} that smooth maps $F:\  \s \to \s$,
which are solutions of  above equation, contain two countable
families of harmonic representatives in the homotopy classes of
degree $0$ and $1$. For a given map the index $k$ of a map is
defined as the number of times the map crosses the line $F=\pi
/2$. The topological charge is unit for odd and null for even
maps. We will denote further these maps by the symbol $\has{k}$.
Solutions of equation $(\ref{eq:main})$ which in the limit
$L\rightarrow \infty$ tend uniformly to $\has{k}$ will be denoted
by $\ha{k}$ and, for convenience, called harmonic maps, though in
fact are not harmonic maps.

To sum up saying \textsl{harmonic} or \textsl{skyrmionic} part of
a solution of equation (\ref{eq:main}) we mean the limiting
behaviour of the solution as $L\rightarrow \infty$ and use these
names for convenience and for the sake of clarity of further
considerations even if $L$ is finite, though for finite $L$ there
is no explicit boundary between harmonic and skyrmionic part. Put
differently in different regions of physical space - the
three-sphere - solutions manifest their different nature. From now
on by saying 'a solution is \textsl{composed} of' a harmonic map
or a skyrmion, we will mean that the solution in the limit of
large $L$ solves approximately - between the poles - the equation
of harmonic maps (\ref{eq:harmap}) and that in the vicinity of
poles (after appropriate rescaling of variables as above) solves
approximately the equation known from the flat Skyrme model
(\ref{eq:skyrmodel}). We shall use above terminology regardless of
a value of the radius $L$.

Because of symmetries of the model we can identify solutions which
can be transformed one into another using translations and
reflections in the target space and reflections in the base space.
From now on we will call such class of equivalence \textsl{a
solution}. Each class has its own representative for which
$F'(0)>0$ and $\q\geqslant 0$ simultaneously, so that by saying
\textsl{a~solution} we mean its representative.  Each solution of
(\ref{eq:main}) we denote by the symbol $\sk{n}\ha{k}\sk{m}$ which
informs us that in the limit of infinite radius of the base space,
the solution takes the following limiting form
$\sks{n}\has{k}\sks{m}$: skyrmions $\sks{n}$ and $\sks{m}$ with
topological charge $\q=n$ or $\q=m$ localized respectively at the
north or the south pole of the base space, and a harmonic map
$\has{k}$ with some index $k$ which connects them. We admit of
$k=0$ and then $\ha{0}$ denotes the neutral solution which is the
representative of trivial solutions $F\equiv k\pi$. Depending on
whether the harmonic map has an odd or an even index, it
contributes or not the unit topological charge
$\q_h=\textrm{sgn}(F'(0))$, where $F$ is a solution of   equation
(\ref{eq:harmap}). Thus the limiting solution
$\sks{n}\has{k}\sks{m}$ has the topological charge $\q=n+\q_h+m$
which of course must be the topological charge of the solution
$\sk{n}\ha{k}\sk{m}$, since topological charge cannot be changed
by continuous deformation within a class of functions fulfilling
given boundary conditions.  By definition for all solutions of
equation (\ref{eq:main}) $n\geqslant 0$ (since $F'(0)>0$) and
$\q\geqslant0$. In special cases $k=m=0$ and, in the limit of
infinite radius, we have a pure skyrmion (localized at the north
pole), or $n=m=0$ and then we have a pure harmonic map
$\has{k}=\lim _{L\rightarrow\infty}\ha{k}$ between three-spheres.
A skyrmionic solution with negative topological charge we call
anti-skyrmion and denote by $\as{}$ \emph{i.e.} if $m<0$ then
$\sk{m}\equiv\as{-m}$. By $\ahs{k}$ we denote the harmonic maps
for which $F'(0)<0$.

Profile functions of solutions of  equation ($\ref{eq:main}$) and
their phase diagrams are shown in this paper using  the conformal
variable $x$ which is related to the angle $\psi$ by the equation
$x=\ln{\tan{(\frac{\psi}{2})}}$.  The profile functions are
plotted using coordinates $(x,F(x))$ and the phase diagrams using
coordinates $(F(x),F'(x))$.

\begin{figure}[h]
\begin{center}
\includegraphics[angle=0,height=0.2\textheight,width=0.8\textwidth]
{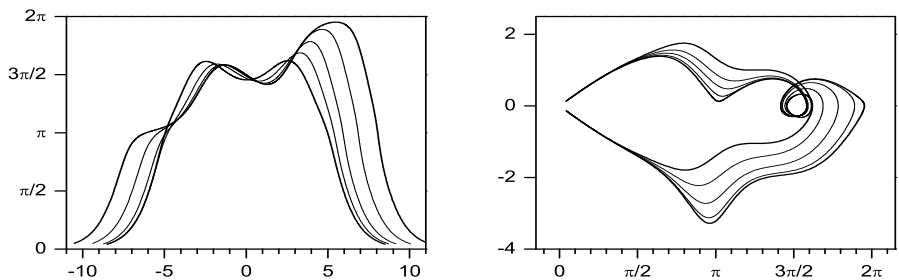} \vspace{-0.02\textheight} \caption{ \small{Solution
$\sk{1}\ha{3}\as{2}$ and its phase diagram for different sizes of
the base space. This solution whose representant is
$\sk{2}\ah{3}\as{1}$ bifurcates from $\sk{1}\ha{4}\as{1}$ at some
critical value of the radius of the base three-sphere.
}}\label{fig:s1h3a2}
\end{center}
\end{figure}

Figure \ref{fig:s1h3a2} illustrates above terminology using as an
example the solution $\sk{1}\ha{3}\as{2}$ together with its phase
diagram for different sizes of the base space . This solution,
whose representant is  $\sk{2}\ah{3}\as{1}$, bifurcates from the
solution $\sk{1}\ha{4}\as{1}$ at the critical value $L\approx
126.56$ of the radius of the base three-sphere. The greater the
radius the more skyrmionic branches of the solution flatten and
become tangent to the lines $F=\pi$ and $F=2\pi$ respectively for
1-skyrmion in the vicinity  of the north pole and for anti
2-skyrmion in the vicinity  of the south pole. The inner part of
the solution in the limit of infinite radius overlaps the harmonic
map $\has{3}$ for which $F(0)=\pi$ and $F(\pi)=2\pi$.

In the following we shall show that the extremely rich structure
of solutions of the nonlinear equation (\ref{eq:main}), undergoes
a strikingly simple scheme which enables us doing predictions of
the properties of the solutions and how do they appear.
\input{howappear}

%% file: howfind.tex
\subsection{Numerics}
The aim of this work is to find the whole spectrum of solutions of
the nonlinear equation $(\ref{eq:main})$. This can be achieved
only by finding numerical solutions. For this purpose it is
essential to know the Taylor series expansion of solutions in the
vicinity of singular points of the equation, namely about $\psi=0$
and $\psi=\pi$. The formal Taylor series expansion for solutions
with $F(0)=0$ is
\begin{eqnarray}F(\psi)=a\psi
-\frac{a(a^2-1)(4+a^2\kappa^2)}{30(1+a^2\kappa^2)}\psi^3+\olarge(\psi^5),
\qquad a=F'(0).\label{eq:expansion}
\end{eqnarray}
This series expansion can be used as a launcher from the singular
point to the domain of regular points of equation (\ref{eq:main})
where numerical integration may be successfully carried on. In the
left neighbourhood of $\psi=\pi$ the series is similar. By the
method of 'shooting to a fitting point' \cite{nr} we are able to
get effectively numerical solutions of  the equation. The series
quite accurately approximates solutions at the boundaries and the
$n+1$ term  can be used to assess an error one does by terminating
the series at $n$'th term. This allows us to move away as far as
possible from the singular points, within the accuracy desired,
and then carry on the integration numerically.

%We take into account only the solutions for which $F(0)$ and
%$F(\pi)$ are some integer multiples of $\pi$. Another possibility
%are the solutions for which $F(0)$ or $F(\pi)$ (or both) are some
%odd multiples of $\pi/2$ (see the appendix for appropriate
%proofs). But then such solutions behave like the function
%$\sqrt{\lambda}\sin {\ln(\lambda)}$ about $\lambda=0$, oscillating
%infinitely many times in the vicinity of poles. From the physical
%reasons we give up considering them, otherwise the energy integral
%($\ref{eq:energint}$) would be divergent.

We denote by $F_o$ the function which we get by carrying on an
integration forward step by step via equation (\ref{eq:main}),
propagating the initial data $F(0)=0$ and $F'(0)=a_o$ from the
north pole of the base three-sphere. Similarly by $F_\pi$ we
denote the function derived by propagating the initial data
$F(\pi)=\q \pi$ and $F'(\pi)=a_\pi$ backward from the south pole.
By assumption $a_o$ is positive. The numbers $a_o$ and $a_\pi$ we
call the \textsl{shooting parameters}.

Within a given topological sector with topological charge
$\q\geqslant0$ and for fixed nonzero radius $L$, there exist some
finite number of  pairs $\{a_o,a_{\pi}\}$ of critical shooting
parameters, such that the functions $F_o$ and $F_\pi$ and their
first derivatives, launched from the opposite poles, will meet at
some fitting point $\psi_f$ which, from the symmetry, is taken to
be the equator $\psi_f=\half \pi$. In this way, for these
exceptional pairs $\{a_o,a_\pi\}$, we get two complementary parts
of the same integral curve which can be matched together to form
the global solution of equation (\ref{eq:main}) with the initial
data $(0,a_o)$ at $\psi=0$ or equivalently $(\q \pi,a_\pi)$ at
$\psi=\pi$. In special cases $a_\pi=a_o$ or $a_\pi=-a_o<0$.

In generic situation if we start numerical integration  with
$F(0)=0$ and with an arbitrary shooting parameter $a_o$, then
there will exist an integer $l$ such that the numerical solution
$F_o(\psi)$, continued up to $\psi=\pi$, will attain the value
$F_o(\pi)=(\half+l)\pi $. In this way we get the function
$y(a_o)=F_o(\pi)$ which is antisymmetric with respect to $a_o$ and
can be restricted to the set $a_o\in \mathbb{R}^+$. The function
takes values in the set $\mathcal{X}:=\{x\in \mathbb{R}:x=(
\half+l)\pi,l\in \mathbb{Z}\}$, apart from some discrete number of
critical values of $a_o$ for which the function attains the values
being some integer multiples of $\pi$. Since $y$ is a step
function, if $a_o$ is critical then the difference
$y(a_o+\eta)-y(a_o)=\pm\pi/2$, for sufficiently small $\eta$, is
positive or negative in dependence on the value of $a_o$, and for
fixed $a_o$ on the sign of $\eta$. Having found numerically the
function $y(a_o)$ at some fixed radius $L$, we are able to find
all critical shooting parameters $a_o$. Of course this can be done
only approximately since, numerically, one can generate the
function $y(a_o)$ only for a discrete number of values of $a_o$.
It should be clear that among the (approximately) critical values
there may exist pairs which are (approximate) shooting parameters
of the same global solution. In particular, if a global solution
has a reflection symmetry in the base space, its pair of critical
shooting parameters may consist of $\{a_o,a_o\}$ or
$\{a_o,-a_o\}$. Next the pairs of approximate shooting parameters
can be used as trial ones to find exact values for $\{a_o,a_\pi\}$
and afterwards to find the global numerical solutions of equation
(\ref{eq:main}). It is achieved by zeroing a norm of the vector
function $(F_o-F_{\pi},F'_{o}-F'_{\pi})$ of two variables $a_o$
and $a_{\pi}$, taken at the fitting point $\psi_f=\pi/2$. For this
purpose it was used the globally convergent Newton-Raphson method
for nonlinear systems of equations \cite{nr}.

%% file: howappear.tex
\section{How do the solutions appear?} In this section we assume
that the parameters $\e$ and $\f$ of the model are finite, nonzero
and fixed once and for all. Using this parameters we define the
units of energy $\f\e^{-1}/2$ and of length $\f^{-1}\e^{-1}$. Then
$L=\f\e R$ has the interpretation of a dimensionless radius of the
base three-sphere. I emphasize the trivial fact the 'coupling
constant' $\kappa^2=2/L^2$ of the model can be changed both by
changing physical radius $R$ or by changing $\e$ or $\f$. In
particular the limit $\kappa^2=0$ can be attained either by taking
$\e\rightarrow \infty $, then we recover the sigma model with the
unit of energy $R\f^2/2$ (energies of harmonic maps are then
finite and the same as in \cite{biz}); or by flattening the
physical space putting $R\rightarrow \infty$, then we recover the
flat Skyrme model with the unit of energy $\f \e^{-1}/2$ (and then
the energies of skyrmionic solutions coincide with those from the
flat Skyrme model). As we shall see this fact partly explains why
the Skyrme model on three-sphere admits existence of branches of
solutions of harmonic, skyrmionic or mixed type, which meet at
some critical value of $\kappa^2$ and disappear.

\subsection{Solutions which exist for all values of the coupling constant}
\input{eternalons}\label{chp:eternalons}
\subsection{Harmonic maps and accompanying solutions}
\input{harmonicons}\label{chp:harmonicons}
\subsection{The n-skyrmions and their companions}\label{chp:skyrmions}
\input{skyrmions}

%% file: eternalons.tex
\begin{figure}[h]
\begin{center}
\includegraphics[angle=0,height=0.4\textheight,width=0.8\textwidth]{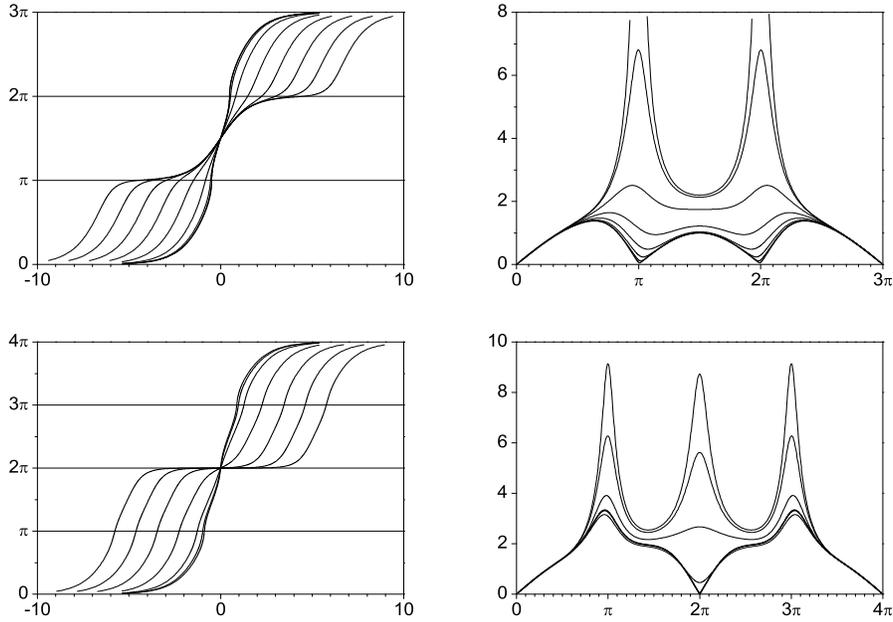}
\vspace{-0.02\textheight} \caption{
\small{Solutions $\sk{1}\ha{1}\sk{1}$  ($\kappa^2=10^{-5},\ldots ,10^3$) and %%@
$\sk{2}\ha{0}\sk{2}$ ($\kappa^2=10^{-3},\ldots ,10^2$) and their phase diagrams for %%@
different sizes of the base space. The smaller the radius $L$ the
higher the maxima for $F'$. In the limit $L\rightarrow 0$ the
maxima become infinite and then the solutions tend to the
solutions of   the limiting equation (\ref{eq:eternalons}). }}
\label{fig:snhsn}
\end{center}
\end{figure}
\begin{figure}[h]
\centering
\includegraphics[angle=0,height=0.3\textheight,width=0.6\textwidth]{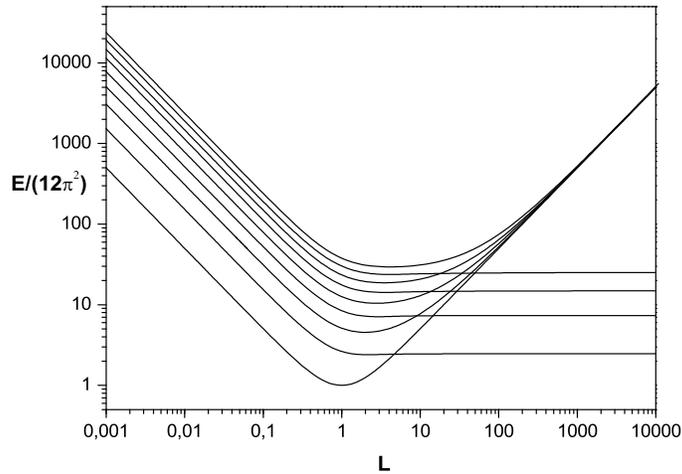}
\vspace{-0.03\textheight} \caption{ \small{Energies (in units
$\f\e^{-1}/2$) for solutions  $\sk{n}\ha{0}\sk{n}$ or
$\sk{n}\ha{1}\sk{n}$ for $n\geqslant 1$. Counting from the bottom
the curves are $\ha{1}$, $\sk{1}\ha{0}\sk{1}$,
$\sk{1}\ha{1}\sk{1}$, $\sk{2}\sk{2}$ and so on with growing
topological charge $\q$ up to $\sk{4}\ha{1}\sk{4}$. It should be
evident how this curves will behave for $\q>9$. The lowest line -
the identity $\ha{1}$ - can be explicitly calculated from equation
(\ref{eq:energid}) hence we get the asymptotic behaviour of
energies for other solutions. }} \label{fig:snhsnen}
\end{figure}
\begin{figure}[h]
\centering
\includegraphics[angle=0,height=0.3\textheight,width=0.6\textwidth]{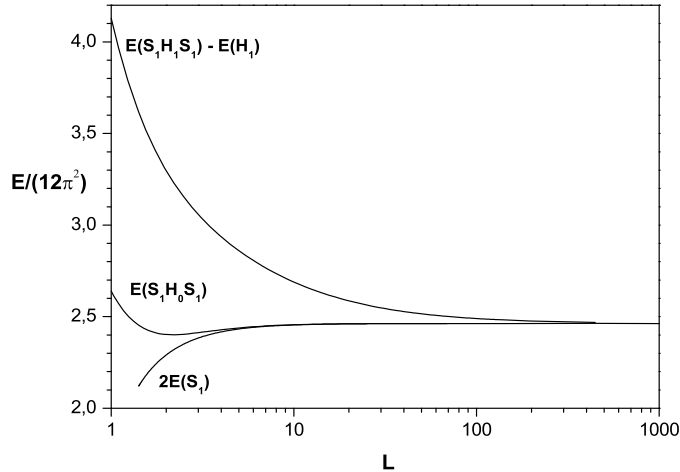}
\vspace{-0.03\textheight} \caption{ \small{This figure shows how
the hypothesis of tripartite structure of solutions works in
practice. The lower line shows the energy (multiplied by $2$) of
1-skyrmion $\sk{1}$ which, having bifurcated from the identity
solution $\ha{1}$ at $L=\sqrt{2}$, evolves becoming true
1-skyrmion (with energy $\approx1.23145\cdot 12\pi^2$) localized
at one of poles  in the limit of infinite radius of physical space
- the three-sphere. The other lines are: the energy for the
solution $\sk{1}\ha{0}\sk{1}$ composed of two 1-skyrmions, and the
energy of the solution $\sk{1}\ha{1}\sk{1}$ (which contains
harmonic map $\ha{1}$ inside and 1-skyrmions attached to its
endings) after subtracting the energy of the harmonic map
$\ha{1}$.
 }} \label{fig:energcomp}
\end{figure}

For all $\kappa^2$ there exist solutions which in the limit
$\kappa^2\to 0$ tend to configurations
$\ha{0},\ \sk{1}\ha{0}\sk{1},\ \sk{2}\ha{0}\sk{2},\ ...,\
\sk{n}\ha{0}\sk{n},\ ...$, with topological charge $\q=2n$ and solutions
which in this limit tend to configurations %%@
$\ha{1},\ \sk{1}\ha{1}\sk{1},\ \sk{2}\ha{1}\sk{2}, \ ...,\ \sk{n}\ha{1}\sk{n}, \ ...$, %%@
with topological charge $\q=2n+1$. They all are invariant under
the transformation $F(x)\rightarrow \q\pi-F(-x)$ which for integer
$\q$ is a symmetry of the equation of motion (and in general is
broken by generic solutions). As is seen on phase diagrams
(fig.\ref{fig:snhsn})
for small $L$ there is no qualitative difference between solutions
$\sk{n}\ha{1}\sk{n}$ and $\sk{n}\ha{0}\sk{n}$ except for the
number of maxima for the function $F'$. However, in the limit of
infinite $L$, they are quite different - $\sk{n}\ha{1}\sk{n}$
contain additionally the harmonic map $\ha{1}$. This is reflected
especially in the asymptotic behaviour of energies of solutions
with odd and even topological charge in the two limits
(fig.\ref{fig:snhsnen}).
In the limit $\f\rightarrow 0$ (no sigma
term)  $F(\psi)$ solves the equation
\begin{equation}F''+F'^2\cot{F}-\frac{1}{2}\frac{\sin{2F}}{\sin^2{\psi}}=0
\label{eq:eternalons}\end{equation} which can be formally obtained
by taking the limit $\kappa^2\rightarrow \infty$ in equation
(\ref{eq:main}). As is seen on phase diagrams
(fig.\ref{fig:snhsn}), in the limit $L\rightarrow 0$ the maxima of
$F'(\psi)$, localized at such $\psi_k$ for which $F(\psi_k)=k\pi$
where $k$ is some positive integer, tend to infinity. This follows
from the formal Taylor series expansion of solutions of equation
(\ref{eq:eternalons}) whose singular part for $F'$ in the
neighbourhood of $\psi_{k}$ is
$$F'(\psi)\sim \frac{(k\pi)^3}{48}|\psi -\psi_k|^{-1/2}.$$
In the limit $L\rightarrow0$ the shooting parameters for the
solutions tend to finite values e.g. $2.5635$, $4.1047$, $5.641$,
$7.1755$, $8.7089$ respectively for $\sk{1}\ha{1}\sk{1}$,
$\sk{2}\ha{1}\sk{2}$, $\sk{3}\ha{1}\sk{3}$, $\sk{4}\ha{1}\sk{4}$,
$\sk{5}\ha{1}\sk{5}$. These values form a sequence which
asymptotically is arithmetic.  The same (but with different
values) holds for solutions $\sk{n}\ha{0}\sk{n}$. For $L$ small
enough the energies $E_{\q}$ of solutions $\sk{n}\ha{1}\sk{n}$ and
$\sk{n}\ha{0}\sk{n}$ behave like $\e_{\q}/L$ where constants
$\e_{\q}$ form an ascending sequence in raising topological charge
$\q$ and asymptotically scale with $\q^2$ that is
$$\lim_{L\rightarrow 0}LE_{\q}(L)=\e_{\q},
\quad \textrm{and} \quad \lim \limits_{\q\rightarrow
\infty}\frac{\e_{\q}}{\q^2}=const.>0.$$ These energies would
become finite again if, instead of $\f\e^{-1}/2$ as the unit of
energy, we chose $R^{-1}\e^{-2}$ for $R$ and $\e$ fixed (but then
we would get quite different model). In the limit of large $L$ the
asymptotic behaviour of energies for solutions
$\sk{n}\ha{0}\sk{n}$ is very different from the behaviour of
$\sk{n}\ha{1}\sk{n}$. This difference comes from the fact (which
is the numerical observation) that energies of n-skyrmions are
bounded while the energies of harmonic maps are unbounded in the
limit $L\rightarrow \infty$ (with $\f\e^{-1}/2$ as the unit of
energy). Due to the tripartite structure of solutions and  the
fact the integral of energy is additive, this behaviour is clear.
This fact (clarified on fig.\ref{fig:energcomp})
is quite general, that is if we take a solution
$\sk{n}\ha{k}\sk{m}$ then
$$\lim \limits_{L\rightarrow \infty}\big(E_{\sk{n}\ha{k}\sk{m}}(L)-
E_{\ha{k}}(L)\big)=\lim \limits_{L\rightarrow
\infty}E_{\sk{n}}(L)+\lim \limits_{L\rightarrow
\infty}E_{\sk{m}}(L)$$ and
$$\lim \limits_{L\rightarrow
\infty} \frac{ E_{\sk{n}\ha{k}\sk{m}}(L)}{E_{\ha{k}}(L)}=1 .$$
From the tripartite structure it is also clear, that in the limit
$L\rightarrow \infty $, the shooting parameters for solutions
$\sk{n}\ha{0}\sk{n}$ and $\sk{n}\ha{1}\sk{n}$ divided by $L$ are
the same as for n-skyrmions (see pt. \ref{chp:skyrmions}).

We can assume that the arms of solutions $\sk{n}\ha{0}\sk{n}$ and
$\sk{n}\ha{1}\sk{n}$ pass from the regime described by equation
(\ref{eq:eternalons}) to the regime when they are skyrmionic arms
(the n-skyrmions), if their phase diagrams flatten at $x=0$.

%% file: harmonicons.tex
\begin{figure}[h]
\begin{center}
\includegraphics[angle=0,height=0.4\textheight,width=0.8\textwidth]
{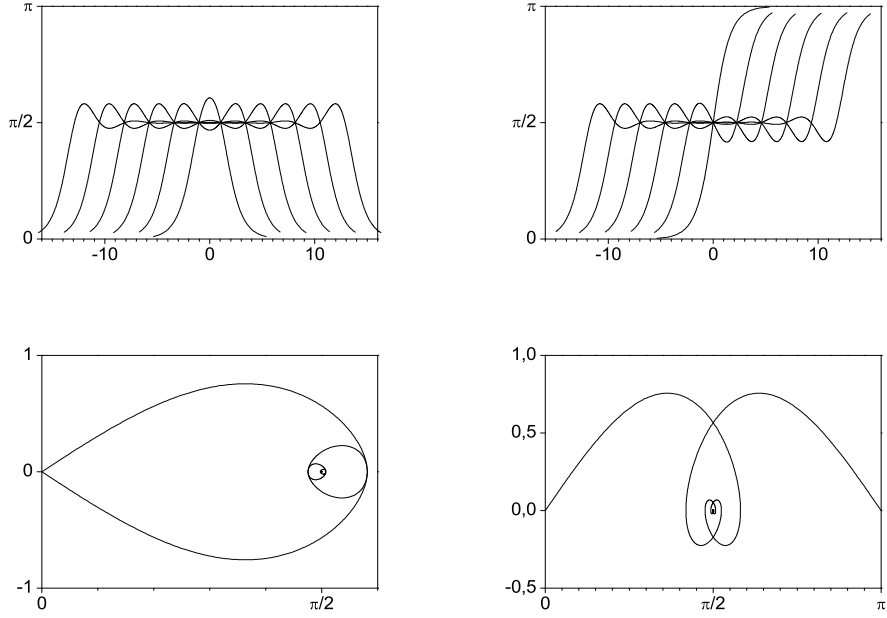} \vspace{-0.02\textheight}
\caption{\small{Even and odd harmonic maps $\ha{k}$ for $k=2,4,6,8,10,12$ and %%@
for  $k=1,3,5,7,9,11$ respectively and phase diagrams for $k=12$ and %%@
$k=11$ in the limit of infinite $\e$. In the %%@
limit $k\rightarrow \infty$} phase diagrams have infinite self-similar %%@
structure.}\label{fig:phaseharm}
\end{center}
\end{figure}
\begin{figure}[h]
\centering
\includegraphics[angle=0,height=0.3\textheight,width=0.6\textwidth]{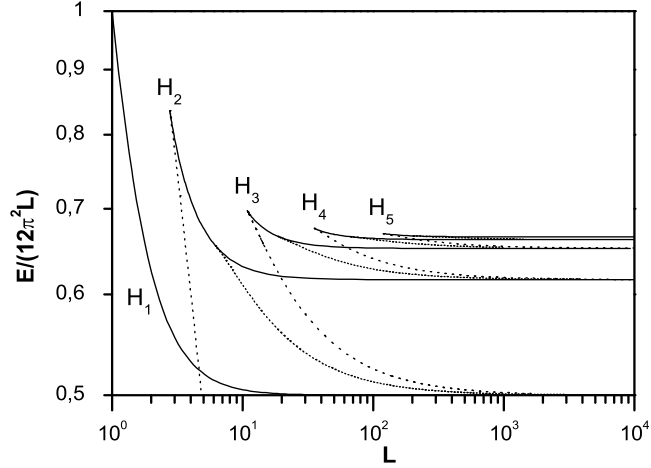}
\vspace{-0.02\textheight} \caption{\small{Energies (in units where
$R\f^2/2=1$) for the first five harmonic maps and the associated
solutions. Solid lines shows harmonic branches $\ha{k}$ (only five
for clarity), dotted lines - branches of associated solutions
$\sk{1}\ah{k-2}\sk{1}$ and $\sk{1}\ah{k-2}\as{1}$ for odd and even
$k\geqslant2$, respectively. These branches coalesce in a cusp for
descending $L$. The short dotted lines are branches of solutions
$\sk{1}\ah{k-1}$ bifurcating from harmonic maps $\ha{k}$. The
identity map $\ha{1}$ is exceptional and exists for all $L$. The
solution $\sk{1}$ bifurcates from $\ha{1}$ at $L=\sqrt{2}$.
}}\label{fig:harmen}
\end{figure}
\begin{figure}[h]
\begin{center}
\includegraphics[angle=0,width=0.6\textwidth,height=0.3\textheight]{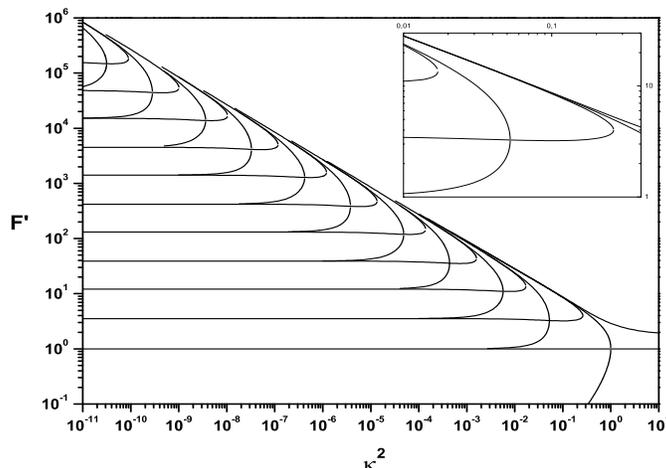}
\vspace{-0.04\textheight} \caption{\small{This figure, which is
asymptotically self-similar, shows branches of solutions for
families of harmonic maps. The lines which asymptotically
($\kappa^2\rightarrow 0$) become horizontal are branches of
tangents $F'(0)$ for the harmonic maps $\ha{k}$ at $\psi=0$.
Tangents $F'(\pi)$ at $\psi=\pi$ are equal to $F'(0)$ and $-F'(0)$
respectively for odd and even harmonic maps. Asymptotic values of
$F'(0)$ are precisely the same as for harmonic maps. If $\kappa^2$
increases then each harmonic branch $\ha{k}$ turns back at a
critical value $\kappa^2_{k}$ and becomes a branch of the
associated solution $\sk{1}\ah{k-2}\as{1}$ for $k$ even and
$\sk{1}\ah{k-2}\sk{1}$ for $k$ odd. All associated harmonic
branches, in the limit $\kappa^2\rightarrow 0$, tend to the common
branch being the line of tangents $F'(0)$ of the 1-skyrmion. The
only exception is the identity $\ha{1}$ which exists for all
$\kappa^2$. It is the first mechanism of how the solutions are
born. To the second mechanism there depend branches crossing the
harmonic ones. The upper branch of the second mechanism is the
line of tangents $F'(0)$ which asymptotically attains the branch
of the 1-skyrmion, while the lower branch is the line of tangents
$F'(\pi)$ or $-F'(\pi)$ if it crosses an odd or an even harmonic
branch. Asymptotically the lower branch is tangent to the harmonic
branch $\ha{k-1}$, so the solutions of the second mechanism have
to be nonsymmetric solutions $\sk{1}\ah{k-1}$. If $k=1$ it is
1-skyrmion alone whose lower branch is asymptotically tangent to
the harmonic branch $\ha{0}$ - the vacuum solution $F=0$.}}
\label{fig:harmpar}
\end{center}
\end{figure}
\begin{figure}[h]
\begin{center}
\includegraphics[angle=0,height=0.4\textheight,width=0.8\textwidth]{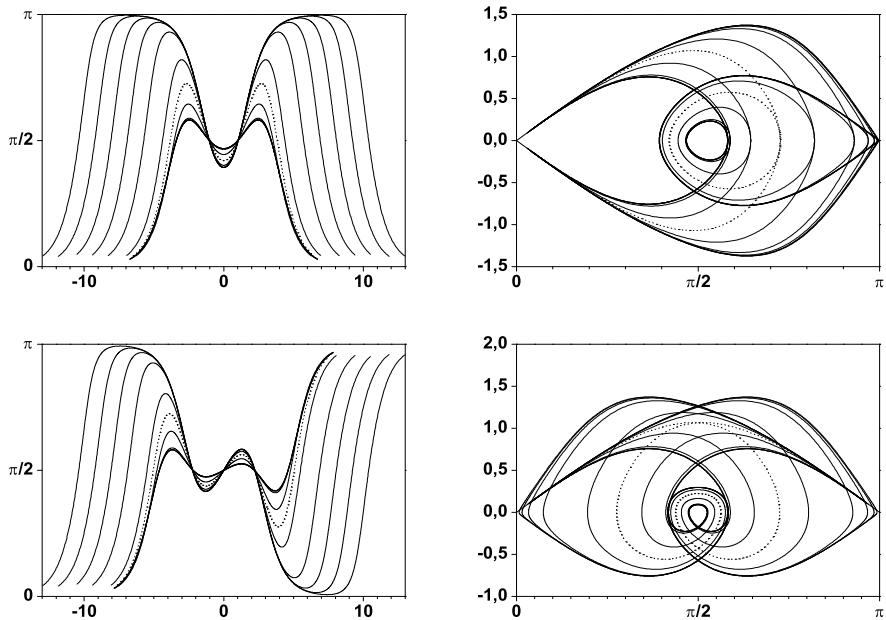}
\vspace{-0.02\textheight} \caption{\small{This figure illustrates
the evolution of even and odd harmonic maps together with
accompanying solutions. The profile functions on the left and
respective phase diagrams on the right are plotted for different
values of coupling constant. The evolution is exemplified by pairs
of solutions: $\ha{4}$ together with $\sk{1}\ah{2}\as{1}$ at the
top and $\ha{5}$ together with $\sk{1}\ah{3}\sk{1}$ at the bottom.
Dotted lines shows the limiting solutions in the vicinity of
critical values of coupling constant $\kappa^2_{\ha{4}}=1.593848
\cdot 10^{-3}$ and $\kappa^2_{\ha{5}}=1.391083 \cdot
10^{-4}$.}}\label{fig:h4h5gr}
\end{center}
\end{figure}
\begin{figure}[h]
\begin{center}
\includegraphics[angle=0,height=0.4\textheight,width=0.8\textwidth]{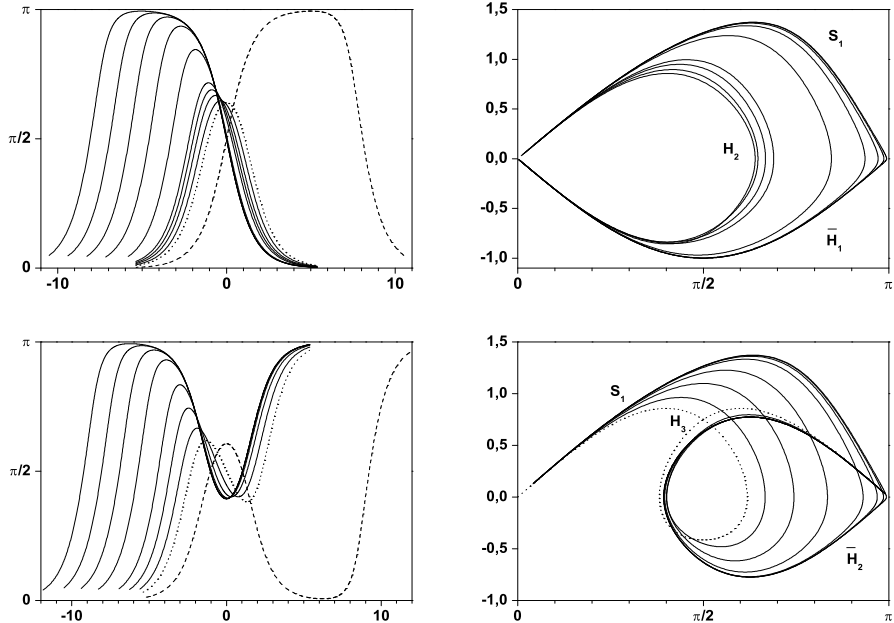}
\vspace{-0.02\textheight} \caption{\small{ The figure shows
profiles and phase diagrams for solutions appearing in the second
mechanism in which new solutions bifurcate from already existing
harmonic maps. Upper figures shows $\sk{1}\ah{1}$ for several
$\kappa^2$ up to $10^{-6}$ (solid lines) and $\ha{1}\as{1}$
(dashed line); both bifurcate from $\ha{2}$ (dotted line) at
$\kappa^2=5.245019\cdot 10^{-2}$. The figures in the bottom show
$\sk{1}\ah{2}$ for several $\kappa^2$ up to $10^{-7}$ (solid
lines) and $\ha{2}\sk{1}$ (dashed line) - both bifurcate from
$\ha{3}$ at $\kappa^2=5.70371 \cdot 10^{-3}$. It is easy to
distinguish skyrmionic and harmonic components of solutions (e.g.
see phase diagrams). In general from harmonic map $\ha{k}$
$k\geqslant1$ there bifurcates $\sk{1}\ah{k-1}$ and simultaneously
$\ha{k-1}\as{1}$ for $k$ even or $\ha{k-1}\sk{1}$ for $k$ odd. In
special case when $k=1$ from harmonic map $\ha{1}$ there
bifurcates 1-skyrmion $\sk{1}$ at the north and at the south pole.
}}\label{fig:h2h3biff}
\end{center}
\end{figure}

There exist a descending sequence of critical values of coupling
constants $\{\kappa_k^2\}_{k \in \mathbb{N}}$, $k\geqslant2$ at
which there appear harmonic maps with index $k$. The only
exception is the identity map $\ha{1}$ which exists for all values
of $\kappa^2$. In the limit $\kappa^2 \rightarrow 0$ all these
maps tend uniformly to those found by \bizon in \cite{biz}. These
limiting maps are shown in figure \ref{fig:phaseharm}.
In the case when this limit is understood in the sense
$\e\rightarrow \infty$ (with $R\f^2/2$ as the unit of energy) the
energies of limiting harmonic maps coincide with those from
\cite{biz}. If the limit $\kappa^2 \rightarrow 0$ was understood
in the sense $R\rightarrow \infty$ (with $\f$ and $\e$ fixed),
then these energies (with $\f \e^{-1}/2$ as the unit of energy)
would be infinite. The energies of harmonic maps and the
accompanying solutions are shown in fig.\ref{fig:harmen}.
The energy of the identity solution $\ha{1}$ can be computed
explicitly from (\ref{eq:energy})
$$E_{\ha{1}}(L)=6\pi ^2(L+L^{-1})\geqslant 12\pi^2.$$ This energy is
infinite in the limit $L\rightarrow \infty$ and attains its
minimum $12\pi^2$ at $L=1$ which is the global minimum of energies
as follows from the Bogomolnyi bound (\ref{eq:bog}). Numerics
shows that in this limit the energies of harmonic maps scale with
$L$ and that the functions  $E_{\ha{k}}(L)/(12\pi^2L)$ have finite
limits
$$\lim \limits_{L\rightarrow \infty}\frac{E_{\ha{k}}(L)}{12\pi^2L}=
\tilde{E}_k<\frac{2}{3},$$ where the sequence
$\{\tilde{E}_k\}_{k\in \mathbb{N}}$ (in units in which
$R\f^2/2=1$) is ascending and converges to $2/3$. This limiting
value is the energy of the singular map $F\equiv \pi/2$ to which
the sequence of functions $\ha{k}(\psi)$ converges pointwise (but
not uniformly because the singular map does not fulfill the
boundary conditions). In the limit $\e\rightarrow \infty$ the
respective sequence of energies ${\scriptstyle \tilde{E}}_k$ is
precisely the same as in the sigma model on $\s$ examined in
\cite{biz}. Furthermore, as we shall see below, after the above
rescaling $E\rightarrow {\scriptstyle \tilde{E}}=E/L$, the
energies of skyrmionic solutions tend to zero so in the limit of
large $L$ different solutions, with different topological charges,
but containing the same harmonic map inside, have the same
limiting 'energies' ${\scriptstyle \tilde{E}}$. Moreover their
profile functions tend pointwise but not uniformly to those of
harmonic maps they contain. This observation qualitatively
explains the conjecture made by \mbox{Y. Brihaye} and \mbox{C.
Gabriel} in \cite{brihaye}, who found several solutions, that
(using my terminology)
\begin{footnotesize}
\begin{eqnarray*}&\lim \limits_{\kappa \rightarrow
\kappa_{2k}}\tilde{E}_{\ha{2k}}(\kappa)=\lim \limits_{\kappa
\rightarrow \kappa_{2k}}
\tilde{E}_{\sk{1}\ah{2k-2}\as{1}}(\kappa)=\tilde{E}_{\ha{2k}}(\kappa_{2k})
,&\lim \limits_{\kappa \rightarrow
0}\tilde{E}_{\sk{1}\ah{2k}\as{1}}(\kappa)=\lim \limits_{\kappa
\rightarrow
0}\tilde{E}_{\ha{2k}}(\kappa)=\tilde{E}_{\ha{2k}}(0),\\
&\lim \limits_{\kappa \rightarrow
\kappa_{2k+1}}\tilde{E}_{\ha{2k+1}}(\kappa)= \lim \limits_{\kappa
\rightarrow \kappa_{2k+1}}
\tilde{E}_{\sk{1}\ah{2k-1}\sk{1}}(\kappa)=\tilde{E}_{\ha{2k+1}}(\kappa_{2k+1}),
&\lim \limits_{\kappa \rightarrow
0}\tilde{E}_{\sk{1}\ah{2k-1}\sk{1}}(\kappa)=\lim \limits_{\kappa
\rightarrow
0}\tilde{E}_{\ha{2k-1}}(\kappa)=\tilde{E}_{\ha{2k-1}}(0).
\end{eqnarray*}\end{footnotesize}\noindent
The numerical evidence for these limits, which are made here far
more plausible, is shown in fig.\ref{fig:harmen} and
\ref{fig:harmpar}.
\\ \indent It should be clear, that in the limit of large $L$
(understood as large $\e$), the Skyrme term can be treated as a
perturbation which deforms harmonic maps from the sigma model on
$\s$. On the other hand the limit of large $L$ has its
unperturbative regime which is attained by fixing $\e$ and taking
the limit $R\rightarrow \infty$, thus there appear other solutions
accompanying harmonic maps. There exist two types of such
solutions which cannot be derived by perturbation from harmonic
maps.  A solution of the first type appears together with
accompanying harmonic map in a characteristic cusp on the diagram
of energies at critical radius $L$. Next they evolve on  separate
branches. (By a \textsl{branch} we mean the plot of the energy or
of the shooting parameters of a solution as a function of $L$ or
$\kappa^2$). A solution of the second type formes another branch
which bifurcates from the harmonic map's branch elsewhere at a
larger critical radius $L$, but before the next harmonic map with
higher index appears (see fig.\ref{fig:harmpar}). These two
mechanisms of appearing of solutions, namely the creation of a
pair and next the bifurcation of a second solution from the just
created harmonic branch, as we shall see further, is generic and
will be repeated during appearing of other solutions with higher
topological charges. The mechanisms in the case of families of
harmonic maps are described below.
\begin{description}
\item[Mechanism I]Each harmonic map $\ha{k}$, $k=2,3\ldots$ appears together with
an accompanying solution. The pattern of appearance is as follows.
The even harmonic map $\ha{2k}$ appears together with the solution
$\sk{1}\ah{2k-2}\as{1}$ \emph{e.g.} $\ha{2}$ together with
$\sk{1}\ha{0}\as{1}$ while the odd harmonic map $\ha{2k+1}$
appears together with $\sk{1}\ah{2k-1}\sk{1}$ \emph{e.g.} $\ha{3}$
together with $\sk{1}\ah{1}\sk{1}$. See tab.~\ref{tab:tabappearh}
and fig.~\ref{fig:h4h5gr}. \end{description}
\input{tabappearh}
The higher the index of a map, the larger $L$ at which it appears.
Here again, in connection with harmonic maps, as in \cite{biz}
there appears the characteristic number $e^{2\pi / \sqrt{7}}$ but
this time it is the limit of a sequence defined in the following
numerical hypothesis:
\begin{hpt} The sequence
$\{\frac{\kappa_{\scriptscriptstyle{k}}}
{\kappa_{\scriptscriptstyle{k+2}}}\}_{k\in \mathbb{N}}$ of
quotients of critical coupling constants at which there appear odd
or even harmonic maps is decreasing and attains the limiting
value $(\sim 10.749087)$\\
\begin{minipage}[h]{\textwidth}
%\begin{table}[h]
\centering \bigskip
\begin{tabular}{|c|}
\hline \\
\raisebox{8pt}{\begin{Large}$\lim \limits_{\scriptscriptstyle{n\to
\infty}} \frac{\kappa_{k}}{\kappa_{k+2}}= e^{2\pi /
\sqrt{7}}$\end{Large}}
\\ \hline
\end{tabular}
%\end{table}
\end{minipage}
\end{hpt}
Using this hypothesis it is easy to predict when and together with
what solution other harmonic maps will appear. This reveals the
whole structure of appearance of harmonic maps together with their
companions. But that is not the whole story. Numerics shows that
there have to exist some (conformal) instabilities in solutions
containing harmonic maps which lead to new solutions which
bifurcate from the previous. This leads to the existence of the
second mechanism of how solutions are born.
\begin{description}
\item[Mechanism II] from each even harmonic map $\ha{2k}$ there bifurcates the
solution $\sk{1}\ah{2k-1}$ and simultaneously its reflection in
the base space $\ha{2k-1}\as{1}$. Similarly from each odd harmonic
map $\ha{2k+1}$ there bifurcates the solution $\sk{1}\ah{2k}$
together with $\ah{2k}\as{1}$ which is derived from the first by
reflection both in the base and in the target space. See
tab.\ref{tab:tabappearh} and fig.\ref{fig:h2h3biff}.
\end{description}

Here again, as in the case of the first mechanism, one can form
the sequence $\{\frac{\kappa_{\scriptscriptstyle{k}}}
{\kappa_{\scriptscriptstyle{k+2}}}\}_{k\in \mathbb{N}}$ of
quotients of critical coupling constants at which  new pair of
solutions bifurcates from odd or even harmonic maps. Again this
limit attains the magic value $e^{2\pi / \sqrt{7}}$.

It is worth noting that the index and topological charge of
solutions appearing in pairs is the same. Similarly the same is in
the case of an already existing solution and the one which
bifurcates from it. I state this as an empirical fact which is
true for all solutions of the Skyrme model on $\s$ irrespective of
a value of the topological charge, but by an index we mean then
the index of a solution within the $\pi$-wide strip which contains
the harmonic map. Summing up, the formation of solutions in the
second mechanism is governed by the laws of conservation of the
index and of the topological charge of appearing solution.

%% file: tabappearh.tex
\begin{table}[h]
\begin{tabular}{||ll|l|l||ll|l|l||}
\hhline{|t:==:=:=:t:==:=:=:t|}
\multicolumn{4}{||c||}{\textsf{mechanism I}}&
\multicolumn{4}{c||}{\textsf{mechanism II }}\\
\hhline{||----||----||} \multicolumn{2}{||c|}{\textsf{pair
}}&\multicolumn{1}{c|}{$\kappa^2_{k}$}&
\multicolumn{1}{c||}{$\kappa_{k}/\kappa_{k+2}$}&
\multicolumn{2}{c|}{\textsf{pair
}}&\multicolumn{1}{c|}{$\kappa^2_{k}$}&
\multicolumn{1}{c||}{$\kappa_{k}/\kappa_{k+2}$}\\
\hhline{|:==:=:=::==:=:=:|}
%********* h2 *************
$\ha{2}$&$\sk{1}\as{1}$&$2.613327\cdot
10^{-1}$&\multicolumn{1}{c||}{$-$}&$\sk{1}\ah{1}$&
$\ha{1}\as{1}$&$5.245019\cdot 10^{-2}$&\multicolumn{1}{c||}{$-$}\\
%********* h3 *************
$\ha{3}$& $\sk{1}\ah{1}\sk{1}$&$1.706483\cdot
10^{-2}$&\multicolumn{1}{c||}{$-$}&$\sk{1}\ah{2}$&
$\ha{2}\sk{1}$&$5.703710
\cdot 10^{-3}$&\multicolumn{1}{c||}{$-$}\\
%********* h4 *************
$\ha{4}$&$\sk{1}\ah{2}\as{1}$&$1.593848\cdot
10^{-3}$&$12.80482$&$\sk{1}\ah{3}$&$\ha{3}\as{1}$&
$4.372362\cdot 10^{-4}$&$10.95256$\\
%********* h5 *************
$\ha{5}$& $\sk{1}\ah{3}\sk{1}$&$1.391083\cdot
10^{-4}$&$11.07578$&$\sk{1}\ah{4}$& $\ha{4}\sk{1}$&$4.849196
\cdot 10^{-5}$&$10.84536$\\
%********* h6 *************
$\ha{6}$&$\sk{1}\ah{4}\as{1}$&$1.351911\cdot 10^{-5}$&$10.85799
$&$\sk{1}\ah{5}$& $\ha{5}\as{1}$&$3.767585
\cdot 10^{-6}$&$10.77275$\\
%********* h7 *************
$\ha{7}$& $\sk{1}\ah{5}\sk{1}$&$1.197673\cdot
10^{-6}$&$10.77724$&$\sk{1}\ah{6}$&
$\ha{6}\sk{1}$&$4.188957\cdot 10^{-7}$&$10.75925$\\
%********* h8 *************
$\ha{8}$&$\sk{1}\ah{6}\as{1}$&$1.167865\cdot
10^{-7}$&$10.75914$&$\sk{1}\ah{7}$& $\ha{7}\as{1}$&$3.259313
\cdot 10^{-8}$&$10.75149$\\
%********* h9 *************
$\ha{9}$& $\sk{1}\ah{7}\sk{1}$&$1.036053\cdot
10^{-8}$&$10.75172$&$\sk{1}\ah{8}$& $\ha{8}\sk{1}$&$3.624800
\cdot 10^{-9}$&$10.75006$\\
%********* hA *************
$\ha{10}$&$\sk{1}\ah{8}\as{1}$&$1.010587\cdot
10^{-9}$&$10.75002$&$\sk{1}\ah{9}$& $\ha{9}\as{1}$&$2.82075
\cdot 10^{-10}$&$10.74931$\\
%********* hB *************
\hhline{~~~~~~--||} $\ha{11}$& $\sk{1}\ah{9}\sk{1}$&$8.966418\cdot
10^{-11}$&$10.74933$&$\sk{1}\ah{10}$&
$\ha{10}\sk{1}$&\multicolumn{2}{c||}{}\\
%********* hC *************
$\ha{12}$&$\sk{1}\ah{10}\as{1}$&$8.746281\cdot
10^{-12}$&$10.74918$&$\sk{1}\ah{11}$&
$\ha{11}\as{1}$&\multicolumn{2}{c||}{\small\textsf{{too poor numerical accuracy}}}\\
%********* hD *************
$\ha{13}$& $\sk{1}\ah{11}\sk{1}$&$7.760219\cdot
10^{-13}$&$10.74911$&$\sk{1}\ah{12}$&
$\ha{12}\sk{1}$&\multicolumn{2}{c||}{}\\
\hhline{|b:==:=:=:b:==:=:=:b|}
\end{tabular}
\medskip
\caption{ Solutions accompanying harmonic maps appear in two
mechanisms. In the first mechanism there appear even harmonic maps
$\ha{2k}$ together with $\sk{1}\ah{2k-2}\as{1}$ or odd harmonic
maps $\ha{2k+1}$ together with $\sk{1}\ah{2k-1}\sk{1}$. In the
second mechanism new solutions bifurcate from already existing
harmonic maps \emph{i.e.} $\sk{1}\ah{2k-1}$, simultaneously with
$\ha{2k-1}\as{1}$, bifurcate from $\ha{2k}$ while $\sk{1}\ah{2k}$,
simultaneously with $\ha{2k}\sk{1}$, bifurcate from $\ha{2k+1}$.}
\label{tab:tabappearh}
\end{table}

%% file: skyrmions.tex
\begin{figure}[h]
\centering
\includegraphics[angle=0,height=0.3\textheight,width=0.6\textwidth]{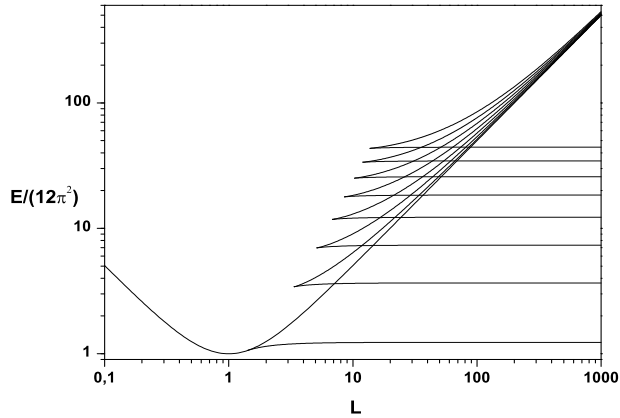}
\vspace{-0.04\textheight} \caption{ \small{The energies (in units
$\f\e^{-1}/2$) of solutions $\sk{n}$ and $\sk{n-1}\ha{1}$ for
$n=1,2,\ldots 8$. The asymptotically horizontal lines are energies
of skyrmionic solutions $\sk{n}$ which, in the limit of infinite
radius $L$, agrees with the energies of skyrmions from the Skyrme
model on flat space. The other lines, which join the branches of
$\sk{n}$ in cusps, are the energies of the solutions
$\sk{n-1}\ha{1}$. The only exception is the solution $\ha{1}$
which exist for all $L$. For $L$ large enough the energy of
$\ha{1}$ behaves as $6\pi^2L$. The same asymptotic behaviour holds
for $\sk{n-1}\ha{1}$. The larger  topological charge $\q=n$, the
larger the radius at which $\sk{n}$ appears. }}
\label{fig:snenerg}
\end{figure}
\begin{figure}[h]
\begin{center}
\includegraphics[angle=0,height=0.25\textheight,width=1\textwidth]{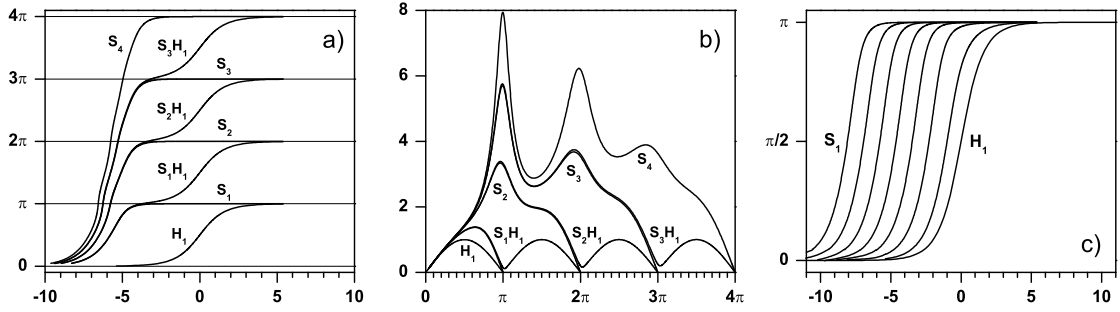}
\vspace{-0.06\textheight} \caption{\small{ (a) The first four
pairs of solutions $\sk{n+1}$ and $\sk{n}\ha{1}$ for
$n=0,\ldots,3$ at $L=100\sqrt{2}$ and (b) their phase diagrams. At
this $L$ there already exist skyrmions with topological charge up
to $82$! (c) The 1-skyrmion $\sk{1}$ bifurcates from the harmonic
map $\ha{1}$ at $L=\sqrt{2}$ and evolves to its limiting
configuration - 1-skyrmion from flat space; the evolution is shown
for radii up to $L=1000\sqrt{2}$ which form geometrical sequence.}
}\label{fig:harm}
\end{center}
\end{figure}

The solutions whose shooting parameters in the limit of large $L$
have asymptotic behaviour of the type $a_0\sim \olarge(L)$ and
simultaneously $a_{\pi}\sim \olarge(L^{-2})$, where $a_o$ and
$a_{\pi}$ are the values of $F'(0)$ and $F'(\pi)$ respectively, we
refer to as n-skyrmions and denote them by $\sk{n}$ where $n$ is
topological charge. The name \textsl{skyrmion} for these solutions
is justified by the observation that in the limit of infinite
radius $L$ of the base three-sphere their energies are just the
same as in ordinary flat Skyrme model (see tab.\ref{tab:asymptsn}
and fig.\ref{fig:snenerg}).
\input{tabasymptsn}
As we have already seen, after appropriate rescaling of
independent variable, in the limit $L\rightarrow \infty$ the
skyrmionic solutions fulfil the Skyrme equation
(\ref{eq:skyrmodel}) in flat space. These solutions we order with
increasing topological charge which simultaneously is the order in
which they appear as the radius $L$ increases. A solution
$\sk{n}$, for which $n\geqslant2$, appears together with its
companion $\sk{n-1}\ha{1}$ which is composed of the harmonic map
$\ha{1}$ and attached to it \mbox{($n-1$)-skyrmion} (see
fig.\ref{fig:harm}).
In the special case when $n=1$ the scheme of appearance is quite
different - the solution $\sk{1}$ bifurcates from the already
existing $\ha{1}$ at the critical radius $L_1=\sqrt{2}$. The
table~\ref{tab:appearsn}
\input{tabappearsn}
shows the sequence of critical radii $\{L_n\}_{n\geqslant2}$ and
respective accompanying solutions.

\begin{figure}[h]
\centering
\includegraphics[angle=0,height=0.3\textheight,width=0.6\textwidth]{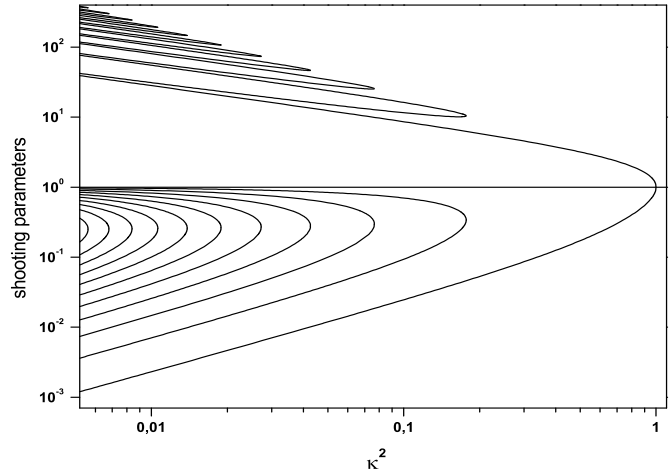}
\vspace{-0.03\textheight} \caption{ \small{Shooting parameters
$a_o=F'(0)$ and $a_{\pi}=F'(\pi)$ for solutions $\sk{n}$ and
$\sk{n-1}\ha{1}$, $n=1,2,\ldots 11$. For the identity solution
$\ha{1}$ $a_o=a_{\pi}=1$ (the horizontal line) for all $\kappa^2$.
For other solutions $a_o>1>a_{\pi}$. For all $n\geqslant2$ the
branches of $a_{\pi}$ and $a_o$ of solutions $\sk{n}$ and
$\sk{n-1}\ha{1}$ meet at critical value $\kappa^2_n$ and
disappear, besides for all $n$ $\kappa^2_{n}<\kappa^2_{n-1}$ and
for all $\kappa^2<\kappa^2_n:$ $a_o(\sk{n})>a_o(\sk{n-1}\ha{1})$
and $a_{\pi}(\sk{n})<a_{\pi}(\sk{n-1}\ha{1})$. Moreover,
$a_o(\sk{n-1}\ha{1})$ tends asymptotically to $a_o(\sk{n-1})$ and
$a_{\pi}(\sk{n-1}\ha{1})$ tends to $a_{\pi}(\ha{1})\equiv 1$. }}
\label{fig:snh1+sn}
\end{figure}

Numerics shows that critical values $L_n$ scale with topological
charge $\q=n$ and supports the hypothesis that the limit of the
sequence $\{L_n/n\}_{n\in\mathbb{N}}$ exists and is finite
$$\lim\limits_{n\to\infty}\frac{L_n}{n}=s_c.$$
A fit gives the value $s_c\approx 1.7169$. From the table
\ref{tab:asymptsn}. it is clear that shooting parameters $a_{0}$
and $a_{\pi}$ for n-skyrmions $\sk{n}$ scale asymptotically with
the radius $L$ that is the limits $\lim_{L\rightarrow \infty
}a_{0}/L$ and $\lim_{L\rightarrow \infty }a_{\pi}L^2$ exist and
are finite (see fig.\ref{fig:snh1+sn}).
In addition this asymptotic equalities, as well limiting energies,
scale again with topological charge and this all is summed up by
the following limits:
$$\lim\limits_{n\rightarrow\infty}\lim\limits_{L\rightarrow\infty}
\frac{1}{12\pi^2}\frac{E_n(L)}{n^2}=s_e, \quad
\lim\limits_{n\rightarrow\infty}\lim\limits_{L\rightarrow\infty}
\frac{a_{0,n}(L)}{Ln}=s_o, \quad
\lim\limits_{n\rightarrow\infty}\lim\limits_{L\rightarrow\infty}
L^2\frac{a_{\pi,n}(L)}{n^2}=s_{\pi}.$$ Fits give approximate
values to the above limits: $s_e\approx 0.688$, $s_o\approx1.77$,
$s_{\pi}\approx 0.24$. The value of the asymptotic energy of the
skyrmionic solution $\sk{1}$ is the same as the energy of
$1$-skyrmion on flat space and confirms results given in
\cite{manton} where the Skyrme model on $\s$ was investigated.
From the table~\ref{tab:asymptsn}. it is clear that n-skyrmions
with $\q=n\geqslant2$ are not energetically stable because
$E_n/E_1>n$. In particular, $E_2>2E_1$ what suggests  that
dynamically $\sk{2}$ would decay into a system of two 1-skyrmions
(maybe $\sk{1}\ha{0}\sk{1}$?). This suggest that n-skyrmions are
unstable (for instance in the case of flat space skyrmions stable
configurations are not spherically symmetric).

%% file: tabasymptsn.tex
\begin{table}[h]
\begin{tabular}[]{||l||*{8}{l|}|}
\hhline{|t:=:t:*{8}{=:}t|}
&$\sk{1}$&$\sk{2}$&$\sk{3}$&$\sk{4}$&$\sk{5}$&$\sk{6}$&$\sk{7}$&$\sk{8}$\\
\hhline{|:=::*{8}{=:}|}
\dv{\scriptstyle{\lim\limits_{L\rightarrow\infty}E_{n}/(12\pi^2)}}
{\scriptstyle{\lim\limits_{L\rightarrow\infty}E_{n}/(12\pi^2n^2)}}
&\dv{1.23145}{1.23145}&\dv{3.66707}{0.91677}&\dv{7.33917}{0.81546}&
\dv{12.2539}{0.76587}&\dv{18.4135}{0.73654}&\dv{25.8186}{0.71718}&
\dv{34.4701}{0.70347}&\dv{44.3680}{0.69325}\\
\hhline{||-||*{8}{-|}|}
\dv{\scriptstyle{\lim\limits_{L\rightarrow\infty}a_{0,n}/L}}
{\scriptstyle{\lim\limits_{L\rightarrow\infty}a_{0,n}/(Ln)}}
&\dv{2.00754}{2.00754}&\dv{3.93062}{1.96531}&\dv{5.77722}{1.92574}&
\dv{7.59385}{1.89846}&\dv{9.39525}{1.87905}&\dv{11.1877}{1.86462}&
\dv{12.9745}{1.85349}&\dv{14.7572}{1.84466}\\
\hhline{||-||*{8}{-|}|}
\dv{\scriptstyle{\lim\limits_{L\rightarrow\infty}L^2a_{\pi,n}}}
{\scriptstyle{\lim\limits_{L\rightarrow\infty}L^2a_{\pi,n}/n^2}}
&\dv{0.45830}{0.45830}&\dv{1.3707}{0.3427}&\dv{2.7540}{0.3060}&
\dv{4.6096}{0.2881}&\dv{6.9386}{0.2775}&\dv{9.7350}{0.2704}&
\dv{13.0132}{0.26557}&\dv{16.759}{0.2619}\\
\hhline{|b:=:b:*{8}{=:}b|}
\end{tabular}
\medskip\medskip
\caption{Limiting values of energies (with $\f\e^{-1}/2$ as the
unit of energy) and scaling of shooting parameters of skyrmionic
solutions. These limiting values scale again with topological
charge in the limit of large values of topological charge.}
\label{tab:asymptsn}
\end{table}

%% file: tabappearsn.tex
\begin{table}[h]
\begin{tabular}{||ll|l|l||ll|l|l||}
\hhline{|t:==:=:=:t:==:=:=:t|}
 \multicolumn{2}{||c|}{pair}&$L_n$&$L_n/n$&
 \multicolumn{2}{|c|}{pair}&$L_n$&$L_n/n$\\
 \hhline{||==:=:=||==:=:=||}
$\sk{2}$&$\sk{1}\ha{1}$&$3.36368$&$1.68184$&$\sk{7}$&$\sk{6}\ha{1}$&$12.0117$&$1.71596$\\
$\sk{3}$&$\sk{2}\ha{1}$&$5.1125$&$1.70417$&$\sk{8}$&$\sk{7}\ha{1}$&$13.7306$&$1.71633$\\
$\sk{4}$&$\sk{3}\ha{1}$&$6.8442$&$1.71105$&$\sk{9}$&$\sk{8}\ha{1}$&$15.4489$&$1.71654$\\
$\sk{5}$&$\sk{4}\ha{1}$&$8.5694$&$1.71388$&$\sk{10}$&$\sk{9}\ha{1}$&$17.1666$&$1.71666$\\
$\sk{6}$&$\sk{5}\ha{1}$&$10.2914$&$1.71523$&$\sk{11}$&$\sk{10}\ha{1}$&$18.8839$&$1.71672$\\
\hhline{|b:==:=:=:b:==:=:=:b|}
\end{tabular}
\medskip \caption{Skyrmionic solutions $\sk{n}$ with $n\geqslant2$ and
their associative solutions $\sk{n-1}\ha{1}$ appear at critical
values of radius $L_n$. The sequence $\{L_n/n\}_{n\in\mathbb{N}}$
has a finite limit.}\label{tab:appearsn}
\end{table}

%% file: wholestructure.tex
\section{Full structure of solutions of the Skyrme model on $\s$}

The main result of this work, besides the statement of the
tripartite structure of solutions of the Skyrme model on $\s$, is
the discovery in what configurations do the solutions appear. My
analysis of a vast number of numerical solutions of equation
(\ref{eq:main}) and their evolution with the radius of the base
three-sphere, had led to stinkingly simple picture. If a new
solution appears when the radius exceeds its critical value,
characteristic for that solution, then either it appears together
with the second accompanying solution or it bifurcates from an
existing solution. The clue in guesswork was to divide all the
solutions into disjoint families $\c{n,k}{i}$ as shown in the
table \ref{tab:wholeappear}.
\begin{table}[h]
\begin{tabular}{||c||ll|c||ll|c||}
\hhline{|t:=:t:==:=:t:==:=:t|}
$n$&\multicolumn{3}{|c||}{$\c{n,k}{0}$}&
\multicolumn{3}{c||}{$\c{n,k}{1}$}\\
\hhline{||~||---||---||} &\multicolumn{2}{|c|}{\textsf{pairs of
solutions}}&$\q$&
\multicolumn{2}{c|}{\textsf{pairs of solutions}}&$\q$\\
\hhline{|:=::==:=::==:=:|} &$\sk{n}\ah{2k-2}\as{n}
$&$\sk{n-1}\ha{2k}\as{n-1}$&$0$&
$\sk{n}\ah{2k-1}\as{n-1}$&$\sk{n-1}\ha{2k}\as{n-1}\ \star $&$0$  \\
$k>0$&$\sk{n}\ah{2k-2}\as{n-m}$&$\sk{n-1}\ha{2k-1}\as{n-m}$&$m$&
$\sk{n}\ah{2k-1}\as{n-m}$ & $\sk{n-1}\ha{2k}\as{n-m}$&$m-1$\\
&$\sk{n}\ah{2k-2}\sk{n-1}$&$\sk{n-1}\ha{2k-1}\sk{n-1}\
\star$&$2n-1$&
$\sk{n}\ah{2k-1}\sk{n}$&$\sk{n-1}\ha{2k+1}\sk{n-1}$&$2n-1$\\
\hhline{||-||---||---||}$k=0$&
\multicolumn{2}{c|}{$\sk{n}\sk{n}$}&$2n$&
\multicolumn{2}{c|}{$\sk{n}\ha{1}\sk{n}$}&$2n+1$\\
\hhline{|b:=:b:==:=:b:==:=:b|}
\end{tabular}\medskip
\caption{Disjoint families of solutions $\c{n,k}{i}$ of the Skyrme
model on $\s$ where $n\in \mathbb{N}\cup\{0\}$, $i=0,1$ (we say
that $i=0$ is opposite to $i=1$). Solutions $\sk{n}\sk{n}$ and
$\sk{n}\ha{1}\sk{n}$ which exist for all radii $L$ of base
three-sphere are included conventionally in the families. The
solutions from $\c{n,k}{i}$ marked by the star $\star$ belong to
families with subscript $n-1$ and with the opposite $i$ and are
unstable at certain critical value of $L$ when they 'emit' a
solution written on its left side. The positive integer $m$ takes
the values $1,2,\ldots, 2n-2$ if $i=0$ or $2,3,\ldots,2n-1$ if
$i=1$.} \label{tab:wholeappear}
\end{table}
The nonnegative index $n$ means that within the family
$\c{n,k}{i}$ we ask how the solution $\sk{n}\ah{p(i,k)}\sk{r}$,
which contains the n-skyrmion at the north pole, harmonic map
$\ah{}$ with index $p$ inside, and r-skyrmion at the south pole;
appears under assumption that $r$ takes such values that $r$ does
not exceed $n$ and that the topological charge of the solution is
not negative. The topological charge depends of course on whether
the solution contains an odd or an even harmonic map. To see how
this all works I give here several examples:
\begin{itemize}
\item
\begin{description}
\item[Solutions $\sk{n}\ha{0}\sk{n}$ and $\sk{n}\ha{1}\sk{n}$] The solutions
    $\sk{n}\ha{0}\sk{n}$ from $\c{n,0}{0}$ never decay in contrast to
  the solutions $\sk{n}\ha{1}\sk{n}$ from $\c{n,0}{1}$ from which $\sk{n+1}\sk{n}$ bifurcate
  belonging to the families $\c{n+1,1}{0}$. As we saw before the
  solutions $\sk{n}\sk{n}$ and $\sk{n}\ha{1}\sk{n}$ exist for all
  radii of the base three-sphere.\end{description}
  \item
    \begin{description}
    \item[Solutions $\ha{k}$ ('harmonic maps')] Families $\c{1,k}{0}$ and $\c{1,k}{1}$ contain
  even and odd harmonic maps respectively. From the table
  \ref{tab:wholeappear}.
  we read off that the maps $\ha{2}$, $\ha{4}$,
  \ldots, appear together with $\sk{1}\as{1}$, $\sk{1}\ah{2}\as{1}$,
  \ldots,
  respectively, and odd maps $\ha{3}$, $\ha{5}$, \ldots, appear respectively
  together with $\sk{1}\ah{1}\sk{1}$, $\sk{1}\ah{3}\sk{1}$,\ldots.
  We saw in \S\ref{chp:harmonicons} that the greater the index of a harmonic map, the greater the
   radius $L$ at which
  it appears. Nevertheless, before $\ha{p+1}$ appears, $\sk{1}\ah{p-1}$ bifurcates
  from $\ha{p}$. Of course, the solutions $\ha{p}$
  for $p=2,3,\ldots$ are not harmonic maps (we have used this name for convenience), but in the limit of
  infinite radius of the base three-sphere they also fulfil the equation of
  harmonic maps between three-spheres.     \end{description}
\item
    \begin{description}
    \item[skyrmionic solutions] Families $\c{n,1}{0}$ for $n=2,3,\ldots$
  contain n-skyrmions $\sk{n}$ which appear together with
  $\sk{n-1}\ha{1}$. We remember from \S\ref{chp:howfind} how
  to interpret the name $n$-skyrmion of the solution $\sk{n}$
  which,
  for finite radii of the base three sphere, is not $n$-skyrmion at all. But
  the similarities and proper interpretation introduced in
  previous paragraphs had justified the terminology.
    \end{description}
\item
    \begin{description}
    \item[Solutions with $\q=0$] contain even harmonic maps which
    can appear only in pairs $\sk{n}\ha{2k}\as{n}$ together with
    $\sk{n+1}\ah{2k-2}\as{n+1}$, whereas solutions containing odd
    harmonic maps $\sk{n}\ah{2k-1}\as{n+1}$ appear from decay of
    $\sk{n}\ha{2k}\as{n}$ where $k\geqslant1$ and $n\geqslant0$.
    Further it turns out that the greater $n$ with fixed $k$ or
    the greater $k$ at fixed $n$, the greater
    critical radius at which given solution appears.
    %But if we took two
    %arbitrary solutions with $\q=0$ and with $\min(n,k)$ arbitrarily
    %large, we could not say which one would appear earlier. It is
    %because of finite resolution of numerical methods.
    \end{description}
\end{itemize}
Numerics shows that in general there is no rule which could be
used to "forecast" the order of appearance of arbitrary solutions
(at least if one uses numerical methods of finding solutions). If
we take the family of solutions with n-skyrmion at the north and
m-skyrmion at the south pole fixed but which are distinguished by
the harmonic maps with different index $k$ they contain, then the
cusps of the type as in fig.\ref{fig:snenerg} or knee-like shapes
as in fig.\ref{fig:harmpar} and in fig.\ref{fig:snh1+sn},
 locate, in the limit of large $k$, along
straight lines (if we use logarithmic scales) whose slopes are
different in dependence on a pair $(n,m)$. In this limit, along
given straight line, these critical values are distributed
homogenously what implies that asymptotically they form a
geometrical sequence with base quotient being certain power of
$10^{2\pi/\sqrt{7}}$. Now it is clear that the sequences of
critical values from different lines are not commensurate with
each other and that is why it is impossible to predict which
solution would appear as the next one after the one arbitrarily
chosen.

\begin{figure}[h]
\begin{center}
\includegraphics[angle=0,height=0.3\textheight,width=0.6\textwidth]{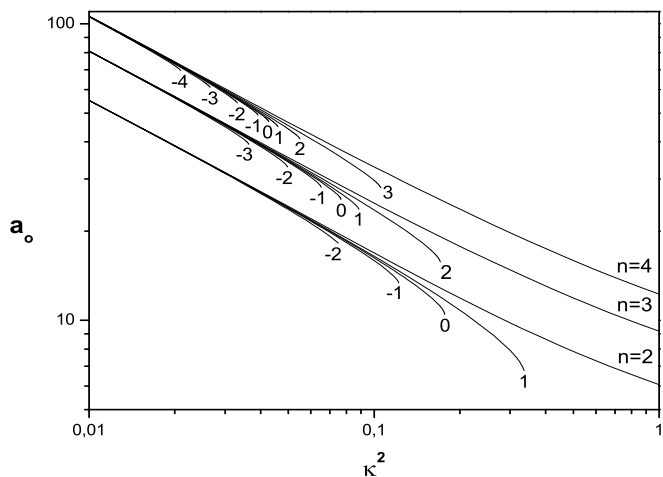}
\vspace{-0.03\textheight} \caption{\small{Families of solutions
from $\c{n,1}{0}$ for $n=2,3,4$. A number at each branch, showing
shooting parameters $a_o$ as functions of $\kappa^2$, is the
topological charge of skyrmionic solution localized at the south
pole. }}\label{fig:sspar}
\end{center}
\end{figure}

In spite of above we can distinguish certain 'directions' in which
such 'putting in order of appearance' is possible. Here are some
of them:
\begin{enumerate}
\item
If we take the set of solutions $\sk{n}\ha{p}\sk{m}$, with
arbitrary but fixed $n$ and $m$, with different $p\geqslant0$,
then the larger $p$, the greater the radius at which the solution
containing harmonic map $\ha{p}$ appears.
\item
For fixed topological charge, the solution
$\sk{n+s}\ha{p}\as{m+s}$ appears at larger radius than the
solution $\sk{n}\ha{p}\as{m}$.
\item For fixed nonnegative $n$, the solution $\sk{n+s}\sk{m}$
appears at larger radius than $\sk{n}\sk{m}$ where $s\geqslant1$
and $0\leqslant n-m<2n$.\label{3}
\item If positive $n$ is fixed then sequentially there appear
$\sk{n}\sk{n-1}$, $\sk{n}\sk{n-2}$, \ldots, $\sk{n}\sk{-n}$ (see
fig.\ref{fig:sspar}). \label{4}
\end{enumerate}
We can include solutions containing harmonic maps into (\ref{3})
and (\ref{4}). Using above rules, reflections in the base and/or
in the target space, and using the table \ref{tab:wholeappear}, we
can generate much more rules. But they are not all of course.
Using above empirical rules we are able to 'predict' for example
that the solution $\sk{2}\sk{1}$ appears earlier (at lower $L$)
than $\sk{2}$; it means that although there exists a solution
(\emph{e.g.} $\sk{2}\sk{1}$) there do not exist all its components
yet (\emph{i.e.} $\sk{2}$ does not exist yet).

%% file: acknwld01.tex
\section*{Acknowledgements}
\medskip

I would like to thank Dr Piotr \bizon, my supervisor, for his many
suggestions and constant support during my research. I am also
grateful for providing me with his private correspondence with Dr
Tadeusz Chmaj which contained first results and ideas being a
source of inspiration to my research.

% ----------------------------------------------------------------------

%% file: appendix.tex
\section{}
\subsection{Local
existence theorem and asymptotics for solutions}
 In section (\ref{chp:skyrmions}) by change of variable $\psi=r/2L$ we
 defined the function $S_L(r)=F(\psi)|_{\psi =r/2L}$ which in the
 limit $L\rightarrow \infty$ fulfils the Skyrme equation
 ($\ref{eq:skyrmodel}$). Solutions of the equation are critical points of
 the energy functional
 \begin{equation}
E[S]=4\pi \int \limits_0^\infty \bigg[\bigg(S'^2+2\frac{\sin
^2{S}}{r^2}\bigg)+\alpha^2\bigg(S'^2+\frac{1}{2}\frac{ \sin
^2{S}}{r^2}\bigg)\frac{\sin ^2{S}}{r^2}\bigg] r^2\ud
r,\label{eq:enfunctapp}
\end{equation}
where $\alpha^2$ depends on a unit of length one chooses (the
second integrand above is not scale invariant) and here
$\alpha^2=8$. For proofs of asymptotic behaviour of solutions of
the equation
\begin{equation}
(\sin ^2{\psi }+\kappa ^2\sin ^2{F})F'' +\sin{2\psi
}F'+\frac{1}{2}\kappa ^2\sin{2F}(F')^2-\sin{2F}-\frac{1}{2} \kappa
^2 \frac{\sin ^2{F}}{\sin ^2{\psi }}\sin{2F}=0,\label{eq:main2}
\end{equation}
the crucial point is the introduction of a function $Q$ which (by
analogy with \cite{mat}) can be simply guessed by comparison of
the functionals (\ref{eq:enfunctapp}) and (\ref{eq:energy}).
\begin{dfn}\label{dfn:q}
\begin{equation}
Q(\psi ):=\frac{1}{4} F'^2(\sin ^2{\psi }+ \kappa ^2 \sin
^2{F})-\frac{1}{2} \sin ^2{F}-\frac{1}{8} \kappa ^2\frac{\sin
^4{F}}{\sin ^2{\psi }},\quad \psi \in [0,\pi]. \label{eq:q}
\end{equation}
\end{dfn}

In this section I exploit methods elaborated by McLeod and Troy in
their work
 \cite{mat} where they gave theorems concerning solutions of the Skyrme
 model on $\mathbb{R}^3$,
  and in principle I repeat their ideas extending them on equation
  (\ref{eq:main2}). In this (numerical) work I need only the proof
  that the series expansion (\ref{eq:expansion}) for solutions of the Skyrme model on $\s$
  for which $F(0)$ and $F(\pi)$ are integer multiples of $\pi$ is valid.
\begin{lem}\label{lem:q}
As $\psi \searrow  0$ the function  $Q(\psi )\sin ^2{\psi }$
converges to a finite limit.
\end{lem}
$\Box .$ Let $\psi \in [0,\half \pi )$ then the function $g=\sin
^2{\psi }(Q+\half )$ is increasing because from the definition of
$Q$
\begin{displaymath}g'=\sin ^2{\psi }(Q'+2Q\cot {\psi }+\cot {\psi })
=4\sin ^2{F}\cot{\psi }(\oct \kappa ^2F'^2-\quarter )+\cot{\psi
}\geqslant \cos ^2{F}\cot {\psi }\geqslant 0.\end{displaymath} So
that $g$ decreases as $\psi \searrow  0$ and the function $Q\sin
^2{\psi }$ has to attain a limit which is finite or not. But  it
cannot equal $-\infty $ what would imply from $Q\sin ^2 {\psi }$
that $\quarter \sin ^2{\psi }F'^2 (\sin ^2{\psi }+\kappa ^2\sin
^2{F})$ was negative what is impossible.  Hence there exists a
constant $a$ such that
\begin{equation}
\lim _{\psi \rightarrow 0^+}Q(\psi )\sin ^2{\psi }=a, \quad
\textrm {and $a$ is finite.} \label{eq:01}
\end{equation} \begin{flushright}$\Box .$\end{flushright}
What is the consequence of this limiting behaviour of $Q(\psi )$
for convergence of $F(\psi )$ at $\psi =0$? The answer is the
following lemma.
\begin{lem}As $\psi \searrow 0$, the function $F(\psi )$ tends to a  limit.
\label{lem:limzero}\end{lem} $\Box .$ For contradiction suppose
that $F(\psi )$ does not tend to a limit as $\psi \searrow  0$,
hence $\sin {F}$ does not tend to a limit too. From (\ref{eq:q})
and (\ref{eq:01})
 it is seen that
$$\quarter F'^2\sin ^2{\psi }(\sin ^2{\psi }+\kappa ^2 \sin ^2{F})
-\half \sin ^2 {\psi }\sin ^2{F}\sim a+\oct \kappa ^2 \sin ^4{F}$$
and next that
$$\quarter \kappa ^2F'^2\sin ^2{\psi }\sin ^2{F}\sim a+\oct \kappa ^2
\sin ^4{F}$$where we used the fact that since $\sin ^2{F}$ does
not tend to a limit then $(\sin ^2{\psi }+\kappa ^2 \sin
^2{F})\sim \kappa ^2 \sin ^2{F}$ and since $\sin {F}$ is bounded
then $-\half \sin ^2{\psi } \sin ^2{F}\rightarrow 0$. Moreover we
can choose a descending sequence of points $\{\psi _n\}$ in such a
way that $\lim_{n\to \infty}\psi _n=0$ and that simultaneously
$\oct \kappa ^2 \sin ^4{F(\psi _n} )\not= -a$ (if $a< 0$) and
$\oct \kappa ^2 \sin ^4{F(\psi _n)}\not= 0$ for all $n$. For these
$\psi _n$  $\quarter \kappa ^2F'^2\sin ^2{\psi }\sin ^2{F}$ is
finite and different from $0$. Having noticed that
$$F'\sin{\psi }=\psi F'
(1+\osmall(\psi))\quad \textrm{if} \quad \psi<\varepsilon, $$ we
see that $\psi _n F'(\psi _n )$ is finite and different from $0$.
In particular there exist such $\Delta$ that $0<\Delta <1$ and for
all $n$ $\psi F'(\psi )$ is bounded and different from zero in
some interval $\psi _n(1-\Delta )\leqslant \psi \leqslant \psi
_n(1+\Delta)$. On the other hand $$(Q(\psi )\sin ^2{\psi })'=\sin
^2{F}\cot {\psi }(\half \kappa ^2F'^2\sin ^2{\psi }-\sin
^2{\psi}),$$ and since $\psi F'$ is finite and bounded from zero
there must exist such constant $M>0$ that for sufficiently large
$n$
$$(Q(\psi )\sin ^2{\psi })'\geqslant \frac{M}{\psi _n}\quad
\textrm{if only }\quad \psi _n(1-\Delta )\leqslant \psi \leqslant
\psi _n(1+\Delta).$$ Integrating this in the interval $\psi
_n^-\leqslant \psi \leqslant \psi _n^+$ where $\psi _n^{\pm }=\psi
_n(1\pm \Delta)$ we get that
$$s(\psi _n^+)-s(\psi _n^-)\geqslant 2M\Delta, \quad \textrm{and}
\quad s(\psi )=Q(\psi )\sin ^2{\psi }.$$ But this says that
$Q(\psi )\sin ^2{\psi }$ as a function does not fulfils the Heine
criterion for convergence, what contradicts the fact proved in
previous lemma that  $Q(\psi )\sin ^2{\psi }$ tends to a limit if
$\psi \searrow  0$.
 Thus $F(\psi )$ tends to a limit if $\psi \searrow  0$.
\begin{flushright}$\Box .$\end{flushright}
\begin{lem}As $\psi \uparrow \pi$, the function $F(\psi )$ tends to a  limit.\end{lem}
$\Box .$  The function $\widetilde{F}(\widetilde{\psi })=F(\psi
),\ \widetilde{\psi }\in [0,\pi ]$ where $\psi =\pi
-\widetilde{\psi }$ fulfils identical equation like $F(\psi )$
that is why all qualitative properties of the function $F(\psi )$
pass to the function $\widetilde{F}(\widetilde {\psi })$. In
particular $\widetilde{F}(\widetilde{\psi })$ tends to a limit as
$\widetilde{\psi }\searrow  0$ what is equivalent to that $F(\psi
)$ tends to a limit as $\psi \uparrow \pi $.
\begin{flushright}$\Box .$\end{flushright}
\begin{thm}
As $\psi \searrow  0$ then every solution tends to a limit which
may equal either $k\pi $ or  $(k+\half )\pi $, where $k\in
\mathbb{Z}$.  As $\psi \uparrow \pi $ then every solution tends to
a limit which may equal $l\pi $ or $(l+\half )\pi $, where $l\in
\mathbb{Z}$.
\end{thm}

\emph{Proof.} Suppose that the limit $\lim _{\psi \rightarrow
0^+}\sin{F(\psi )}$, which exists by virtue of lemma
(\ref{lem:limzero}), is different from zero what means that
$\sin{F_o}\not= 0$ where $F_o=\lim _{\psi \rightarrow 0^+}F(\psi
)$. From definition (\ref{dfn:q})  and from the lemma
(\ref{lem:q}) we get then
\begin{displaymath}\kappa ^2 F'^2 \sin ^2{F}\sin ^2{\psi }\sim
\half \kappa ^2\sin ^4{F_o}+4a,\end{displaymath}
and further that
\begin{displaymath}F'^2\sim \frac{8a+ \kappa ^2\sin ^4{F_o}}{2
\kappa ^2\sin ^2{F_o}\sin ^2{\psi }}\sim \frac{C_1}{\psi
^2}.\end{displaymath} We must put $8a+ \kappa ^2\sin ^4{F_o}=0$
otherwise $F(\psi )$ would not be bounded because then $F(\psi
)=\olarge (\ln{\psi })$ and it  would contradict the lemma
(\ref{lem:limzero}). But then $F'(\psi )=\osmall (\psi ^{-1})$ and
taking it into consideration in equation (\ref{eq:main2})
multiplied by $\sin ^2{\psi }$ and taking the limit $\psi \searrow
0$ we get
\begin{displaymath}\lim _{\psi \rightarrow 0^+}\sin ^2{\psi }
F''(\psi )=\sin{F_o}\cos{F_o}.\end{displaymath} Now if $\cos
{F_o}\not= 0$ ($\sin {F_o}\not= 0$ from assumption) then  $F'(\psi
)=\olarge (\psi ^{-1})$ what cannot be because it contradicts just
now derived formula $F(\psi )=\osmall (\psi ^{-1})$.  Thus if
$\sin {F_o}\not= 0$ then $\cos {F_o}=0$.
 So the conclusion is that the only possible limiting values of
 $F(\psi )$ at zero may be looked for
among the numbers from the countable set $\{k\frac{\pi }{2} : k\in
\mathbb{Z}\}$. As in previous proof we state that the behaviour of
solutions in the left-sided neighbourhood of $\psi =\pi $ is
similar i.e.  the only possible limiting values of $F(\psi )$ at
$\pi $ may be looked for among the numbers from the countable set
$\{l\frac{\pi }{2} : l\in \mathbb{Z}\}$. This completes the proof.

We are interested in behaviour of solutions of   equation
($\ref{eq:main2}$) in the vicinity  of the singular points $\psi
=0$ and $\psi=\pi$ for which $F(\psi)\rightarrow k\pi$ with an
integer $k$. Because of reflection symmetries of the equation we
can limit ourself by considering only the solutions for which
$F(0)=0$ and $F'(0)>0$. For such solutions $F(\psi)>0$ and
$F'(\psi)>0$ for sufficiently small $\psi$. Irrespective of any
behaviour of $F$ as $\psi\rightarrow 0$ we can for contradiction
take a sequence of points $\{\psi_n\}_{n\in\mathbb{N}}$  for which
$\psi_n\rightarrow0$ and $F(\psi_n)>0$, $F'(\psi_n)>0$. If for
large enough $n$ there was $F(\psi)>0$ and $F'(\psi)=0$, where
$\psi>\psi_n$, then from (\ref{eq:main2}) we see that $F''(\psi)$
would be positive and so $F$ would have local minimum at this
$\psi$. Thus for sufficiently small $\psi $ we have always
$F(\psi)>0$ and $F'(\psi)>0$. We shall use this fact further.

The formal expansion (\ref{eq:expansion}) suggests that $\psi
F'/F\rightarrow 1$ as $\psi \rightarrow 0$. This is the hint for
the proof of asymptotics of solutions of the equation
(\ref{eq:main}).

\begin{lem}%\label{lem:q}
As $\psi\searrow 0$ then $f(\psi)<1$ holds for the function
$f(\psi):=\frac{F'(\psi)\sin{2\psi}}{2F(\psi)}$.
\end{lem}
$\Box .$ For contradiction suppose that $F'\sin{2\psi}>2F$ then
$$\bigg(\frac{F'\sin{2\psi}}{2F}\bigg)'=\frac{F''\sin{2\psi}}{2F}+\frac{F'}{2F^2}
(2F\cos{2\psi}-F'\sin{2\psi})<\frac{\sin{2\psi}}{2F}F'',$$ since
from assumption $2F\cos{2\psi}-F'\sin{2\psi}<2F(\cos{2\psi}-1)<0$.
Moreover $$\sin^2{F}-(F')^2\sin^2{\psi}<F^2-\quarter
(F')^2\sin^2{2\psi},$$ so from ($\ref{eq:main2}$) we get $F''<0$.
But then $f(\psi)$ increases as $\psi \searrow 0$ and
$f(\psi)>c>1$ for $\psi $ small enough. In particular
$F'>2cF/\sin{2\psi}$ and integration in the interval
$(\varepsilon_n,\psi_n)$ (where $\varepsilon_n=\psi_n/n$) gives
$$\frac{F(\psi_n)}{\psi_n}>c\frac{2F(\delta_n)}{\sin{2\delta_n}}\left(1-\frac{1}{n}\right)
+\frac{F(\varepsilon_n)}{\varepsilon_n}\frac{1}{n}>c\frac{2F(\delta_n)}{\sin{2\delta_n}}
\left(1-\frac{1}{n}\right),\qquad 0<\varepsilon_n<\delta_n<\psi_n,
$$ and hence, taking the limit $n\rightarrow \infty$, we get $g\geqslant cg$
for $c>1$ where the nonnegative number $g$ is the limit
$g:=\lim_{\psi \rightarrow 0 }F(\psi)/\psi$. But this can be true
only if $g=0$ and then from (\ref{eq:main2}) we get
$$\psi F''=\olarge(1)\bigg(\frac{\sin{2F}}{\psi}-\olarge(1)F'+\half\kappa^2
\frac{\sin{2F}}{\psi}\bigg(\frac{\sin^2{F}}{\sin^2{\psi}}-(F')^2\bigg)\bigg)<-mF'$$
fore some positive $m$. Integration of $F''/F'<-m/\psi$ gives
$$\ln{\frac{F'(\psi)}{F'(\varepsilon)}}<-m\ln{\frac{\psi}{\varepsilon}},
\qquad 0<\varepsilon<\psi.$$ But this implies that as $\varepsilon
\searrow 0$ (at fixed $\psi$) then $F'(\varepsilon)$ diverges to
$+\infty$ what in turn contradicts that $F(\psi)=\osmall(\psi)$.
Thus the assumption $F'\sin{2\psi}>2F$ is false and always, for
sufficiently small $\psi$, $f(\psi)<1$ must hold. This proves the
lemma.
 \begin{flushright}$\Box .$\end{flushright}
 \begin{lem}
As $\psi\searrow 0$ then $f(\psi)>1$ holds for the function
$f(\psi):=\frac{2F'(\psi)\tan{\psi}}{\sin{2F(\psi)}}$.
\end{lem}
$\Box .$ For contradiction suppose that $2F'\tan{\psi}<\sin{2F}$
then
$$\bigg(\frac{2F'\tan{\psi}}{\sin{2F}}\bigg)'=\frac{2\tan{\psi}}{\sin{2F}}F''
+\frac{2F'}{\sin^2{2F}\cos^2{\psi}}
(\sin{2F}-F'\cos{2F}\sin{2\psi})>\frac{2\tan{\psi}}{\sin{2F}}F'',$$
since from assumption
$\sin{2F}-F'\cos{2F}\sin{2\psi}>2F'\tan{\psi}(1-\cos{2F}\cos^2{\psi})>0$.
Moreover
\begin{eqnarray*}&&\sin{2F}-F'\sin{2\psi}>F'(2\tan{\psi}-\sin{2\psi})>0,\\
\textrm{and}\quad &&
\sin^2{F}-(F')^2\sin^2{\psi}>(F')^2(4\tan^2{\psi}-\sin^2{\psi})>0,
\end{eqnarray*} so from ($\ref{eq:main2}$) we get $F''>0$. But then $f(\psi)$
decreases as $\psi \searrow 0$ and in fact $f(\psi)<c<1$ for $\psi
$ small enough. In particular $F'<c\sin{2F}/2\tan{\psi}$ and
integration in the interval $(\varepsilon_n,\psi_n)$ (where
$\varepsilon_n=\psi_n/n$) gives
$$\frac{F(\psi_n)}{\psi_n}<c\frac{\sin{2F(\delta_n)}}{2\tan{\delta_n}}\frac{\psi_n-\varepsilon_n}{\psi_n}
+\frac{F(\varepsilon_n)}{\varepsilon_n}\frac{\varepsilon_n}{\psi_n}<c\frac{\sin{2F(\delta_n)}}{2\tan{\delta_n}}
+\frac{F(\varepsilon_n)}{n\varepsilon_n},\qquad
0<\varepsilon_n<\delta_n<\psi_n, $$ hence, taking the limit
$n\rightarrow \infty $, $g\leqslant cg$ for $0<c<1$ where the
nonnegative number $g$ is the limit $g:=\lim_{\psi \rightarrow 0
}F(\psi)/\psi$. But this can be true only if $g=0$. The same
arguments as in the previous lemma show that $F'(\psi)$ would  be
infinite in the limit $\psi\searrow 0$ what would contradict that
$F(\psi)=\osmall(\psi)$. Thus the assumption
$2F'\tan{\psi}<\sin{2F}$ is false and always, for sufficiently
small $\psi$, $f(\psi)>1$ must hold, what proves the lemma.
\begin{flushright}$\Box .$\end{flushright}
The inequalities we have already proved can be recast in to the
form
$$ \frac{\psi \sin{2F(\psi)}}{2\tan{\psi}}< \psi F'(\psi)<\frac{2\psi F(\psi)}{\sin{2\psi}}$$
and they are valid for sufficiently small $\psi$. The Taylor
series expansion about $\psi=0$ and $F=0$ gives
$$ -\frac{2}{3}F^3-\frac{1}{3}F\psi^2+\osmall(\psi^3)<\psi F'
-F < \frac{2}{3}\psi^2F+\osmall(\psi^3)$$ what implies that
$$F(\psi)=a\psi +\olarge(\psi^3), \quad \textrm{where} \quad
a=F'(0),$$ and this proves (\ref{eq:expansion}). The same of
course, must hold at $\psi=\pi$ (the only difference is that
$F(\pi)=k\pi$ with an integer $k$, which is irrelevant to the
proof).